\documentclass[a4paper,fleqn,usenatbib,useAMS]{mnras}%,referee]{mnras}

\usepackage{graphicx}	% Including figure files
\usepackage{amsmath}	% Advanced maths commands
\usepackage{amssymb}	% Extra maths symbols
\usepackage{multicol}        % Multi-column entries in tables
\usepackage{bm}		% Bold maths symbols, including upright Greek
\usepackage{pdflscape}	% Landscape pages
\hypersetup{draft}
\newcommand{\sme}{SME}
\newcommand{\afe}{$\alpha$}

\usepackage[T1]{fontenc}
\usepackage{ae,aecompl}
\usepackage{txfonts}
\usepackage{gensymb}
\usepackage{amssymb}
\usepackage{url}
\usepackage{txfonts}

\title[HERBS I: Metallicity and alpha enhancement]{HERBS I: Metallicity and alpha enhancement along the Galactic bulge minor axis}
\author[L. Duong et al.]{L. Duong,$^{1}$ M. Asplund,$^{1,2}$\thanks{Contact email: \href{mailto:martin.asplund@anu.edu.au}{martin.asplund@anu.edu.au}} D. M. Nataf,$^{3}$ K. C. Freeman,$^{1}$ M. Ness$^{4,5}$ and
\newauthor{L. M. Howes$^{6}$}
\\
$^{1}$Research School of Astronomy \& Astrophysics, Australian National University, ACT 2611, Australia\\
$^{2}$ARC Centre of Excellence for All Sky Astrophysics in 3 Dimensions (ASTRO 3D)\\
$^{3}$Center for Astrophysical Sciences and Department of Physics and Astronomy, The Johns Hopkins University, Baltimore, MD 21218, USA\\
$^{4}$Department of Astronomy, Columbia University, Pupin Physics Laboratories, New York, NY 10027, USA\\
$^{5}$Center for Computational Astrophysics, Flatiron Institute, 162 Fifth Avenue, New York, NY 10010, USA\\
$^{6}$Lund Observatory, Department of Astronomy and Theoretical Physics, Box 43, SE-221 00 Lund, Sweden
}

\date{Accepted XXX. Received YYY; in original form ZZZ}

\pubyear{2018}

\begin{document}
\label{firstpage}
\pagerange{\pageref{firstpage}--\pageref{lastpage}}
\maketitle

% Abstract of the paper
\begin{abstract}
\noindent To better understand the origin and evolution of the Milky Way bulge, we have conducted a survey of bulge red giant branch and clump stars using the HERMES spectrograph on the Anglo-Australian Telescope. We targeted ARGOS survey stars with pre-determined bulge memberships, covering the full metallicity distribution function. The spectra have signal-to-noise ratios comparable to, and were analysed using the same methods as the GALAH survey. In this work we present the survey design, stellar parameters, distribution of metallicity and alpha-element abundances along the minor bulge axis at latitudes $b=-10^{\circ}, -7.5^{\circ}$ and $-5^{\circ}$. Our analysis of ARGOS stars indicates that the centroids of ARGOS metallicity components should be located $\approx$0.09 dex closer together. The vertical distribution of $\alpha$-element abundances is consistent with the varying contributions of the different metallicity components. Closer to the plane, alpha abundance ratios are lower as the metal-rich population dominates. At higher latitudes, the alpha abundance ratios increase as the number of metal-poor stars increases. However, we find that the trend of alpha-enrichment with respect to metallicity is independent of latitude. Comparison of our results with those of GALAH DR2 revealed that for [Fe/H] $\approx -0.8$, the bulge shares the same abundance trend as the high-$\alpha$ disk population. However, the metal-poor bulge population ([Fe/H] $\lesssim -0.8$) show enhanced alpha abundance ratios compared to the disk/halo. These observations point to fairly rapid chemical evolution in the bulge, and that the metal-poor bulge population does not share the same similarity with the disk as the more metal-rich populations. 
\end{abstract}

% Select between one and six entries from the list of approved keywords.
% Don't make up new ones.

%%%%%%%%%%%%%%%%%%%%%%%%%%%%%%%%%%%%%%%%%%%%%%%%%%

%%%%%%%%%%%%%%%%% BODY OF PAPER %%%%%%%%%%%%%%%%%%

\section{Introduction}

Despite its prominent role in the formation and evolution of the Galaxy, the bulge is perhaps the least understood stellar population. The bulge is host to a great diversity of stars, with up to five peaks in its metallicity distribution function~\citep{Ness2013,Bensby2017}, including some of the oldest stars in the Galaxy~(see e.g.,~\citealt{Howes2015}; and \citealt{Nataf2016,Barbuy2018} for a review). It is also a major Galactic component, comprising 30\% of the Milky Way's total mass~\citep{Portail2015,Bland-Hawthorn2016}. Studies of the bulge are therefore essential for understanding the formation and evolution of the Milky Way, and by inference, other spiral galaxies. 

Galaxy bulges are typically referred to either as a `classical' or `pseudo'-bulge~\citep{Kormendy2004}. Classical bulges are thought to have formed via rapid dissipative collapse consistent with $\lambda$CDM cosmological predictions~\citep{White1978,Tumlinson2010,Rahimi2010}. The properties of classical bulges largely mirror that of elliptical galaxies: they consist of old stars with random stellar motions. On the other hand, pseudobulges, formed via secular evolution are flatter in shape, contain younger stars and show evidence of cylindrical rotation~\citep{Kormendy2004}. While most nearby galaxies appear to have a pseudobulge, some contain both types of bulges~\citep{Fisher2011,Erwin2015}.  
The traditional view of the Galactic bulge is that it is exclusively old ($>$10 Gyr), based on the observed colour-magnitude diagram in multiple fields (e.g., \citealt{Zoccali2003,Clarkson2008}). Using \textit{Hubble Space Telescope} photometry,~\cite{Clarkson2011} and ~\cite{Bernard2018} estimated the young stellar population ($<$5 Gyr) in the bulge to be $<$3.4\% and 11\%, respectively\footnote{Despite the overall small fraction of young stars in the bulge,~\cite{Bernard2018} found a significant fraction (30-40\%) of super-solar metallicity stars to be younger than 5 Gyr. \cite{Bensby2013} and \cite{Haywood2016} also discussed the possibility of young stars masquerading as an old turn off in colour-magnitude diagrams due to a lack of metallicity information in photometric studies.}. Early detailed abundance studies of the bulge corroborated this view: bulge giants are typically overabundant in \afe-elements such as O, Si and Ti, but especially so in Mg (e.g.,~\citealt{McWilliam1994, Zoccali2006, Lecureur2007}). \cite{Fulbright2007} suggested that the abundances of bulge stars plateau at [Mg/Fe] = 0.3 dex even at super-solar metallicity, and the bulge has a separate chemical enrichment to the disk in the solar neighbourhood. Furthermore, multiple authors found a vertical metallicity gradient in the bulge~\citep{Minniti1995,Zoccali2008}. Together these results were interpreted as signatures of a classical bulge population, formed early and rapidly via mergers or dissipational collapse prior to the formation of the disk (e.g.,~\citealt{Matteucci1990}).

The discovery of a significant fraction of young ($<$5 Gyr), relatively metal-rich ([Fe/H] $>$ $-$0.4) microlensed bulge turn-off and subgiant stars thus came as a surprise~\citep{Bensby2013,Bensby2017}\footnote{\cite{Barbuy2018} argued that due to large uncertainties in the distances of microlensed dwarfs, at least some of these young stars are not part of the bulge, but foreground disk stars.}. The presence of such stars would be inconsistent with the classical scenario and instead point to disk-instabilities channelling stars from the disk into the bulge (e.g.,~\citealt{Athanassoula2005,Martinez-Valpuesta2013,Matteo2014,Fragkoudi2018}). Abundance studies now suggest that the Milky Way bulge and thick disk share strong chemical similarities.~\cite{Melendez2008} found that the \afe-abundance trends in the bulge follow that of the local thick disk. \cite{Alves-Brito2010} reached the same conclusion from their re-analysis of \cite{Fulbright2007}; as have many recent studies~\citep{Gonzalez2011,Johnson2014,Ryde2016,Bensby2017}. However,~\cite{Bensby2017} observed that their microlensed bulge stars lie in the upper envelope of the thick disk, implying that the bulge may have experienced a faster chemical enrichment than typical thick disk stars in the solar neighbourhood. An increasing number of kinematic studies show that the bulge rotates cylindrically~\citep{Kunder2012,Ness2013a,Zoccali2014,Ness2016a,Molaeinezhad2016}, which is also evidence against a primarily classical bulge population. Furthermore, infra-red imaging reveal that the bulge is `boxy', or X-shaped~\citep{Dwek1995,Ness2016b}, and the split red clump observed in photometric studies is often attributed to this X-structure \citep{McWilliam2010,Wegg2013,Nataf2015,Gonzalez2015b,Zasowski2016,Ciambur2017}\footnote{See~\cite{Lee2015,Joo2017} for a different interpretation of the double red clump in relation to the X-shaped morphology of the Galactic bulge. {In addition,~\cite{Lopez-Corredoira2016,Lopez-Corredoira2017} observed an absence of the X-structure in the young, main sequence bulge population and Mira variables.}}. The most recent observational evidence thus point to a primarily pseudobulge population in the Milky Way.  

The metallicity distribution function (MDF) of the bulge has proven to be complex, that there is not yet a consensus on the metallicity range (see \citealt{Barbuy2018} for a review). Many studies have shown that it is composed of multiple components (e.g.,~\citealt{Babusiaux2010,Hill2011,Ness2013,Zoccali2014,Rojas-Arriagada2017,Perez2018}). In particular, \cite{Ness2013} showed that there are up to five components based on $\approx$14 000 bulge red-giant stars. They associated stars with [Fe/H] $\approx$ $-$0.5 with the inner thick disk, while the more metal-rich populations with mean [Fe/H] = $-$0.2 and $+0.2$ differ in their kinematics such that stars with the highest metallicity are more prominent near the plane.~\cite{Ness2013} concluded that these metal-rich populations originated in different parts of the early thin disk due to bar-induced disk instabilities. The strength of each MDF component vary with latitude, manifesting as the vertical metallicity gradient seen in earlier studies. Thus, the vertical metallicity gradient cannot be interpreted as a signature of merger or dissipative collapse bulge formation~\citep{Zoccali2008,Babusiaux2010,Ness2013,Rojas-Arriagada2017}. In agreement with \cite{Ness2013}, \cite{Bensby2017} found five peaks in their MDF for a much smaller sample of microlensed bulge dwarfs and subgiants, four of which matched the ARGOS MDF peaks. The sample age distribution of microlensed bulge stars also show multiple peaks that could be interpreted as star formation episodes in the bulge. \cite{Bensby2017} suggested that the peaks at 11 and 8 Gyr could be the onset of the thick and thin disks; and at 6 and 3 Gyr could be associated with the younger parts of the thin disk/Galactic bar.

It is possible that a classical bulge component exists in the Milky Way despite mounting evidence for a predominantly pseudobulge population. Studies have shown that fields at latitudes $|b|>$5 have a combination of X-shaped and classical bulge orbits~\citep{Ness2012,Uttenthaler2012,Pietrukowicz2015}. In addition, the most metal-poor bulge RR Lyrae stars do not show characteristics of the boxy bulge, such as cylindrical rotation~\citep{Dekany2013,Kunder2016}. Spatial and kinematic results from the GIBS survey~\citep{Zoccali2017} indicate that the metal-poor population of the bulge is {centrally concentrated} and rotates more slowly than the metal-rich population, although the authors do not argue strongly for a classical component. If such a component did exist, disentangling it from those originated in the disk may be very challenging~\citep{Saha2015}. \cite{Schiavon2017} have shown that chemical abundances can serve as a powerful diagnostic for identifying sub-populations in the bulge, having found possible evidence of a dissolved globular clusters using APOGEE abundances. 

Studies of the bulge have previously been hindered by high extinction in the bulge region, and the faintness of bulge stars. The sample sizes are typically small if observed at high resolving power (e.g., \citealt{Johnson2014,Jonsson2017,Bensby2017}). While alpha abundance trends are well established for bulge stars with results from the GIBS, \emph{Gaia}-ESO and APOGEE surveys~\citep{Gonzalez2015a,Rojas-Arriagada2017,Schultheis2017}, information on other elements, especially the neutron-capture elements, are still scarce~\citep{Johnson2012,Swaelmen2016}. In this paper, we present the HERMES Bulge Survey (HERBS), which was designed to be in synergy with the GALAH survey~\citep{DeSilva2015}. Here we aim to provide a large chemical inventory for stars in the bulge by leveraging the wavelength coverage of the HERMES spectrograph, which allows us to obtain chemical abundances for up to 28 elements, including the light, alpha, iron-peak and heavy elements. In addition, we will be using similar spectroscopic analysis method and linelist to the GALAH survey, which facilitates a consistent comparison of the chemical properties of bulge and disk stars. 

\section{Data description}
\subsection{Target selection}

For our observations, we selected giants and red clump stars from the analysed sample of the ARGOS survey~\citep{Freeman2013,Ness2013}. ARGOS stars were selected to be between magnitude $K=11.5$--14 from the 2MASS catalogue~\citep{Skrutskie2006}, with $J,K$ magnitude errors $<$0.06 and all quality flags = 0 \citep{Freeman2013}. To exclude most dwarfs, a colour cut in $(J-K)_0=0.38$ was made; each of the ARGOS field was de-reddened using the \cite{Schlegel1998} reddening map. The magnitude and colour selection of ARGOS aimed to minimise very cool and metal-rich giants, but at the same time include very metal-poor giants \citep{Freeman2013}. Any remaining foreground dwarfs are excluded after the ARGOS stellar parameters analysis based on their surface gravity. 

In order to exclude background and foreground giants, distances were used to infer |$R_\textrm{GC}$| for each star. \cite{Ness2013} computed stellar distances by assuming that stars between $\log (g) = $ 1.8--3.2 and $T_\mathrm{eff}=$ 4500--5300 K are clump giants, and have absolute magnitude M$_K=-1.61 \pm 0.22$ \citep{Alves2000}. For stars that are not located near the clump, M$_K$ is obtained by matching stellar parameters with the closest point on a grid of 10 Gyr BaSTI isochrones. The error of red clump based distances is $\approx$15\%, and $\approx$38\% for isochrone based distances. \cite{Ness2013} noted that their red clump sample could be contaminated with non-clump giants, but this contamination should be small. Furthermore, a small subset of their sample shows that isochrone only and red clump only distances return consistent results \citep{Ness2013}. The ARGOS study defined the bulge region to be within Galactocentric radius |$R_\textrm{GC}| \leq 3.5$ kpc. 

This study aims to obtain a thorough chemical inventory of red clump and giant stars, probing the different sub-populations found by \citep{Ness2013} and their variation with latitude. We have therefore made use of the ARGOS $R_\textrm{GC}$ and [Fe/H] measurements to select stars that most likely reside in the bulge region, i.e. those with $|R_\mathrm{GC}| \leq 3.5$ kpc, and gave greater weights to more metal-poor/metal-rich stars in the selection process. We achieved this by allocating $\approx$100\% of ARGOS stars at low and high metallicity, and $\approx$50\% else where. This ensures that we cover the entire metallicity range and all sub-populations in each field, especially increasing the relative fraction of metal-poor stars. 

\begin{figure}
	\centering
	\includegraphics[width=1\columnwidth]{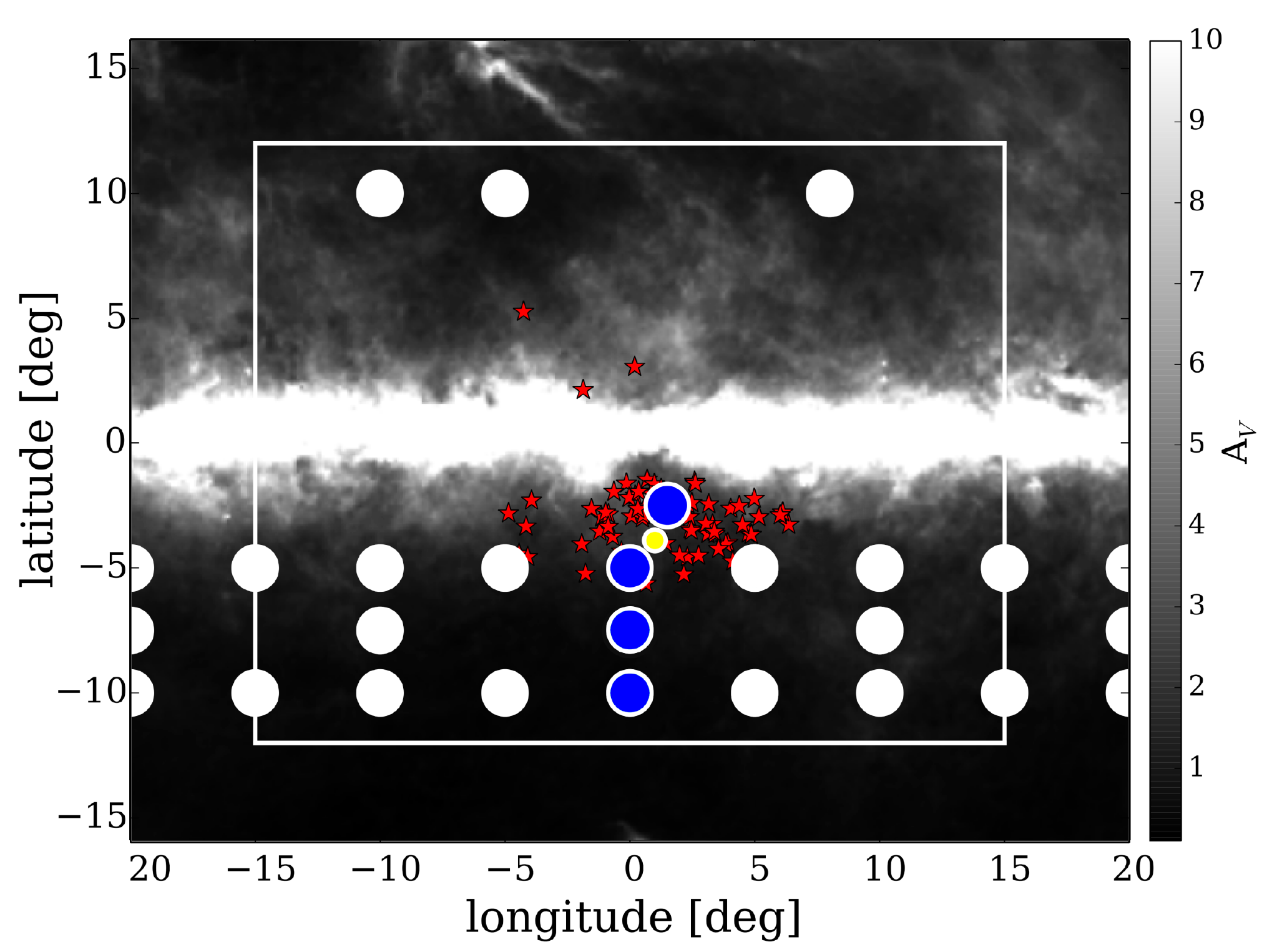}
	\caption{Locations of bulge fields observed by the HERBS survey, shown in blue circles. Also shown are ARGOS fields (white circles, \citealt{Ness2013}), microlensed stars (red stars, \citealt{Bensby2017}) and Baade's Window (yellow circle). The white square indicates roughly the bulge vicinity, and the background map shows the dust extinction.}
    \label{bulgefields}
\end{figure}

Because the integration time required for faint bulge stars is much greater for HERMES than AAOmega (see the next section for details), we could only observe a few ARGOS fields to complete the project in a feasible time frame. Fig.~\ref{bulgefields} shows locations of the observed fields (shaded blue), which includes three ARGOS fields along the minor axis at $(\ell,b)=(0,5)$; $(\ell,b)=(0,-7.5)$; $(\ell,b)=(0,-10)$. In addition to the ARGOS fields, we observed the field $(\ell,b)=(2,-3)$, which was selected due to its relative low extinction and it being covered by K2, which could in principle provide accurate age estimates from asteroseismology (e.g., \citealt{Aguirre2015}). We also added suitable bulge metal-poor candidates ([Fe/H]$_\mathrm{EMBLA} < -1.5$) from the EMBLA survey~\citep{Howes2016} to fields (0,$-10$) and (0,$-5$).

\subsection{Observations}

The observations were taken using the HERMES spectrograph~\citep{Sheinis2015} on the 2dF system of the Anglo-Australian Telescope. The pilot survey, which targeted field $(0,-7.5)$ was completed in August 2014, and observations of the remaining fields were completed between May 2015 and June 2016. 

The 2dF system contains two observing plates that cover a 2$^{\circ}$ diameter field of view. Each 2dF plate has a set of 400 optical fibres, each of them two arcseconds in diameter. Of these, eight fibres are dedicated to bright guide stars to maintain field position accuracy, 25 fibres are allocated to measuring sky variation across the field and typically $\approx$350--360 science objects were observed per field. Sky locations were chosen by visually inspecting DSS images of each field for blank regions. The instrument, HERMES (High Efficiency and Resolution Multi-Element Spectrograph), enables spectra of four wavelength intervals to be observed simultaneously: 4713--4903~\AA~(blue CCD); 5648--5873~\AA~(green CCD); 6478--6737~\AA~(red CCD) and 7585--7887~\AA~(IR CCD). The wavelength coverage of HERMES has been optimised for accurate stellar parameters and abundance measurements, including Balmer lines in the blue and red CCDs. At the nominal resolution of $\lambda/\Delta \lambda \approx 28 000$, HERMES can deliver abundances for up to 28 elements, including Li, O, Na, Mg, Al, Si, K, Ca, Sc, Ti, V, Cr, Mn, Co, Ni, Cu, Zn, Rb, Sr, Y, Zr, Ru, Ba, La, Sm, Ce, Nd and Eu. Combined with the high multiplexity of 2dF/AAT, HERMES is a powerful tool for detail abundance studies of Galactic stellar populations. 

While the optical wavelength coverage of HERMES provides a large number of abundances and accurate parameters, it is also a draw back for bulge observations. Due to the faintness of bulge stars and high extinction in this region, we require significantly longer integration times compared to, for example ARGOS and APOGEE, to achieve the required signal-to-noise ratio (SNR) for precise parameters and abundances. As stated in the previous section, ARGOS stars have 2MASS $K$ magnitude 11.5--14, or an approximate $V$ magnitude of 15--17 in the field $(\ell,b)=(0,-7.5)$. In contrast, typical GALAH targets have $V$ magnitude of 12--14. We aimed to have the same data quality as the GALAH survey, which attains median SNR $\approx$ 100 per resolution element, or $\approx$50 per pixel, for the green CCD \citep{Martell2017}. To determine the required integration time, we observed field $(\ell,b)=(0,-7.5)$ over three consecutive nights and determined the signal-to-noise of co-added spectra in real time. This field has an apparent red clump magnitude of $V \approx 16$, and after 10 hours of observing time we reached the desired signal-to-noise. We scaled this time to estimate the required observing time of all other fields based on their apparent red clump magnitudes. 

The long integration times meant that we must re-configure each field throughout the night to maintain position accuracy. To do this efficiently, we allocated the same set of stars to both 2dF plates, and alternated between them. Observing intervals are split into 30-minutes exposures to minimise the effect of cosmic rays. Calibration frames (fibre flats and ThXe arc frames) were taken either immediately before or after each exposure. Most of the observations were carried out in dark time, some during grey time. Lastly, due to the faint signals of our targets, we chose to observe in the \verb|NORMAL| CCD read-out mode, to minimise read noise while maintaining reasonable overhead time.  

Due to the large fraction of time lost (because of poor weather) over the course of this project, we were not able to complete the observations of fields (0,$-5$) and (2,$-3$). The (0,$-5$) field is lacking some 10 hours, and (2,$-3$) requires approximately 25 additional hours. For this reason, we do not include the $(\ell,b) = (2,-3)$ field in our analysis as the signal-to-noise of this field would be insufficient to derive accurate stellar parameters and abundances. 

The integration time and median SNR in the blue, green and red CCDs of the minor axis fields is given in Table \ref{table:snr}. We were able to achieve similar signal-to-noise to the GALAH survey for the pilot field at $(\ell,b) = (0,-7.5)$ and $(0,-10)$. The median SNR for field $b=-5^\circ$ is much lower, because we were not able to complete the planned observations for this field.

\begin{table}
	\caption{The estimated $V$-magnitude and median signal-to-noise ratio of each bulge field, {except the (2,$-3$) field, which was not used in subsequent analysis due to low SNR}. The IR arm is not shown as it has similar SNR to the red arm. {For HERMES, one resolution element is equivalent to approximately four pixels.}}
	\label{table:snr}
	\begin{tabular}{llllll}
		\hline 
		Field & RC   & Exp time & SNR$_\mathrm{B}$ & SNR$_\mathrm{G}$& SNR$_\mathrm{R}$ \\
		($\ell,b$)  &$V$\textsubscript{mag} & (hours) & (pixel$^{-1}$) &  (pixel$^{-1}$) & (pixel$^{-1}$)  \\
		\hline
		(0, $-5$) & 17.4 & 17 &  20 &  34 &  46\\
		(0,$-7.5$) & 16.3 & 10 & 32 &  51 &  65\\
		(0, $-10$) & 16.0 & 08 & 30 &  40 &  53\\
		\hline
	\end{tabular}
\end{table}

\section{Data reduction}
\label{redux}
Each 30 minute observing block returns a data frame consisting of $\approx$380 spectra (including sky fibres). The data frames were reduced using the standard 2dF reduction package 2dfdr v6.46\footnote{\url{www.aao.gov.au/science/software/2dfdr}}. The software subtracts bias level using the overscan, performs flat-field corrections, calibrates the wavelength using ThXe arclines and subtracts sky. For sky subtraction, we used the throughput mode, in which we calibrated the fibre throughput using strong skylines in the IR arm of HERMES. The reduced frames were checked by eye for consistency and data quality. Frames with low SNR due to clouds, or very poor seeing ($>2$ arcseconds) were excluded after the reduction stage.

2dfdr outputs the calibrated spectra in 400-apertures images, with additional extensions: the fibre table that matches the fibre number to each object and the variance extension. All frames observed within the same night and plate are averaged, weighted by the variance extension. The flux of each spectrum is given by:

\begin{equation}
\label{comb}
flux = \frac{\sum_{i=1}^{n} \left(f_i \sigma_i^{-2}\right)}{\sum_{i=1}^{n} \sigma_i^{-2}} 
\end{equation}
\noindent Here $f_i$ and $\sigma_i$ are the flux and error of an individual spectrum, and $n$ is the number of spectra to be combined. 

\noindent The corresponding variance of the combined spectrum is given by:
\begin{equation}
\label{errcomb}
variance = \frac{1}{\sum_{i=1}^{n} \sigma_i^{-2}}
\end{equation}

\begin{figure*}
	\centering
	\includegraphics[width=1\textwidth]{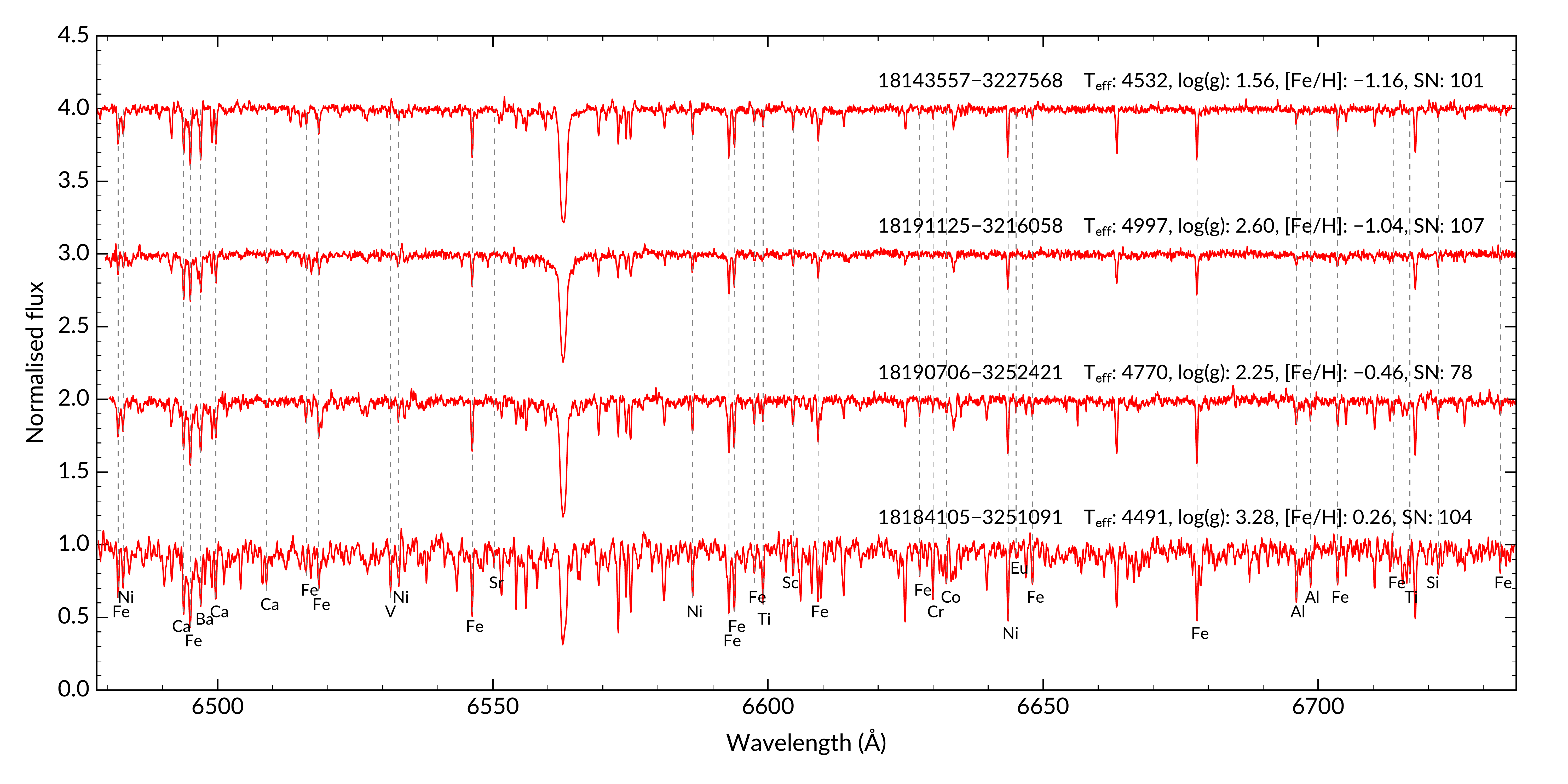}
	\caption{Example HERMES spectra in the `red' wavelength region (which includes the H$\alpha$ line), normalised and shifted to rest using the GUESS code (Section \ref{guess}). The stars' 2MASS identification, SNR (per pixel) and SME-derived parameters are shown. Most of the lines used for stellar parameters and abundance analysis are have been labelled.}
	\label{fig:redspec}
\end{figure*}

To correct for the telluric absorption, we convolved the NOAO atlas~\citep{Hinkle2000} to HERMES spectral resolving power ($\mathcal{R}$ = 28 000). The atlas is scaled to match the typical absorption level at Siding Spring, and shifted by the barycentric velocity of each star. The wavelength points corresponding to telluric lines have their errors increased by the inverse of the telluric absorption level. Spectral pixels affected by tellurics have much lower weights and therefore will not contribute significantly to the spectral synthesis analysis (Section \ref{specanalysis}). Thereafter, each object spectrum is corrected for their barycentric velocity, interpolated onto a common wavelength grid and combined using the same averaging method described above. Examples of reduced spectra can be found in Fig. \ref{fig:redspec}.

We note that our data reduction process is independent of the GALAH reduction pipeline, most importantly we do not correct for the tilted PSF of HERMES spectra~\citep{Kos2017}, which may reduce the resolution and signal to noise towards the corners of each CCD. {The spectral resolving power is lowered by up to $\approx$15\%, and the SNR is lowered by up to $\approx$5\%~\citep{Kos2017}. This affects all fibre bundles, but is minimised towards the CCD centre.}

\section{Radial velocity and initial parameter estimates}
\label{guess}
Having good estimates for the radial velocity and initial parameters significantly speed up the subsequent spectroscopic analysis. For this purpose, we use a modified version of the GUESS code\footnote{\url{https://github.com/jlin0504/GUESS}}, which is implemented in the GALAH survey for radial velocity measurements and has been shown to provide accurate initial parameters (see \citealt{Kos2017}).

Only the blue, green and red HERMES CCDs are used in this step; the IR CCD is excluded as it does not have as many parameter-sensitive lines and is severely affected by telluric absorption. The GUESS code has two separate modules to compute radial velocity and stellar parameters. Radial velocities are calculated via cross-correlation with a grid of 15 AMBRE model spectra~\citep{deLaverny2012}. The models have $\log g = 4.5$, [Fe/H] = 0 and spans 4000--7500 K in $T_\mathrm{eff}$, at 250 K intervals. Prior to cross-correlation, a crude normalisation is done by fitting a spline function over observed spectra, omitting regions around H$\alpha$ and H$\beta$ lines. The normalised spectrum is then cross-correlated with all 15 models, one at a time. The cross-correlation peak is fitted with a quadratic function, the maximum of which is adopted as the cross-correlation coefficient. The coefficients range from 0 to 1, higher values indicate a better match between model and observation. To improve accuracy, model spectra that return coefficients less than 0.3 are excluded. The radial velocity is an average of values from accepted models, weighted by their cross-correlation coefficient. Each HERMES CCD goes through this process independently, and the final radial velocity is an average of all three CCDs, the uncertainty being the standard deviation between CCDs. 

In general our results show good agreement between the three CCDs, with the typical standard deviation being 0.4 km s$^{-1}$. The radial velocity precision also compares well to that reported for the GALAH survey: 93\% of our sample have $\sigma_{v_\mathrm{rad}} \leq 0.6$ km s$^{-1}$, whereas the typical GALAH radial velocity error is $\approx$0.5 km s$^{-1}$~\citep{Zwitter2018}. Our precision may be affected by the lower signal-to-noise ratio of the blue arm in particular, due to the high reddening level in the bulge region.  %The uncertainty is not derived from the peak width of the cross-correlation function because it cannot be calculated accurately from a sample of only 15 cross-correlation functions.

Estimates of stellar parameters ($T_\mathrm{eff}$, $\log g$, [Fe/H]) were derived after the radial velocity determination using a grid of 16783 AMBRE spectra. Spectra are shifted to rest with radial velocities from the previous step, and normalised by fitting 3rd order (for blue and green arms) and 4th order (for the red arm) polynomials to pre-determined continuum regions (by inspection of high resolution spectra of the Sun, Arcturus and $\mu$Leo). The polynomial orders are kept low to avoid poorly constrained continuum fits. From inspecting a large number of normalised spectra, the orders chosen appear to work best for the continuum variation of each CCD. Normalised spectra are interpolated onto the same wavelength grid as the models, and the L$^2$ norm (distance in Euclidean space) between the observed and model spectra are computed. A linear combination of stellar parameters from ten models that are closest in Euclidean space to the observed spectrum give the initial stellar parameters. These were used as starting models in the spectral synthesis analysis described in the next section. 

\section{Spectroscopic analysis}
\label{specanalysis}
\subsection{Stellar parameters}
\label{sp}
The stellar parameters and abundances pipeline and linelist we adopt is the same as that used in Data Release 2 of the GALAH survey~\citep{Buder2018}. The atomic data is based on the $Gaia$-ESO survey linelist~\citep{Heiter2015a}, consisting of mainly blend-free lines with reliable $\log (gf)$ values for stellar parameter determination. However, some background blending lines have slightly different \emph{gf} values compared to the $Gaia$-ESO linelist, as they were changed to improve clearly discrepant fits to the HERMES Arcturus and Solar spectra.
  
For spectral synthesis, we used the code SME (\emph{Spectroscopy Made Easy}) v360~\citep{Valenti1996,Piskunov2017}. In this analysis we implement the 1D, LTE \textsc{{marcs}} model atmospheres~\citep{Gustafsson2008}. The atmospheric models uses spherical geometry with 1$M_\odot$ for $\log g \leq 3.5$, and plane parallel otherwise. During the parameter determination stage, we implement non-LTE corrections from~\cite{Amarsi2016b} for \ion{Fe}{i} lines.

Each spectrum is divided into several $\approx$10~\AA~wide segments containing lines relevant to stellar parameter determination. For this step, there are 20 segments containing line masks for Fe, Ti and Sc. \sme~synthesises the initial model based on the GUESS stellar parameters and radial velocity. In this first iteration, each segment is normalised using a linear function. SME then synthesises lines of H\afe~and H$\beta$; neutral and ionised lines of Sc, Ti, and Fe to determine $T_\textrm{eff}$, $\log g$, [M/H]\footnote{The [M/H], or metallicity parameter is the iron abundance of the best-fit model atmosphere. In our case, this value is very close to the true iron abundance derived from iron lines only. For the purpose of notation consistency when comparing with other studies, we refer to [M/H] as [Fe/H] in subsequent sections.}, $v\sin i$ (rotational velocity) and $v_{\mathrm{rad}}$. {The free SME parameter $v_{\mathrm{rad}}$ is used to bring the model and data spectra to a common wavelength grid. The value of this parameter is typically in line with the radial velocity uncertainty. $v_{\mathrm{rad}}$ is computed independently of other parameters and the same value is used to correct all segments.} 

SME solves for the minimum $\chi^2$ using the Levenberg-Marquardt algorithm. The $\chi^2$ parameter is computed for selected regions following the formula:
\begin{equation}
\chi^2=\frac{\sum\left(\frac{spectrum-model}{variance}\right )^2\times spectrum}{N_\mathrm{lpts}-N_\mathrm{free}-N_\mathrm{seg}}
\end{equation}
\noindent Where $N_\mathrm{lpts}$ is the number of line pixels; $N_\mathrm{free}$ is the number of free parameters and $N_\mathrm{seg}$ is the number of segments. Final parameters from the first cycle are used to build the initial model in the second cycle, which is then used to re-normalise each segment. SME goes through the same iteration process, optimising $\chi^2$ until convergence is achieved (when $\Delta \chi^2 \leq 10^{-3}$). 

Macro-turbulence ($v_{\mathrm{mac}}$) cannot be set as a free parameter for HERMES spectra without causing additional scatter in the results. { This is due to the degeneracy between $v_{\mathrm{mac}}$ and $v\sin i$ at HERMES resolution.} We therefore set all $v_{\mathrm{mac}}$ values to zero, which effectively incorporates $v_{\mathrm{mac}}$ into our $v\sin i$ estimates. Similarly, micro-turbulence $(\xi_t)$ is determined by temperature-dependent formulas that were calibrated for the \emph{Gaia}-ESO survey~\citep{Smiljanic2014}. For giants ($\log g \leq 4.2$) we adopt:
\begin{equation}
\xi_t = 1.1 + 1.0 \times 10^{-4} \times (T_\textrm{eff}-5500)+ 4 \times 10^{-7} \times (T_\textrm{eff}-5500)^2
\end{equation}

The resolving power of HERMES is variable across the CCD image, in both the dispersion and aperture axes. A stable median value can be estimated by interpolating each segment with pre-computed resolution maps from~\cite{Kos2017}; this solution is implemented for the GALAH survey~\citep{Buder2018}. However, since our spectra are combined from different fibres, we cannot recover the resolution information. Thus, for the SME analysis, we adopted $\lambda/\Delta \lambda $ = 28 000 throughout. The synthetic spectra are convolved with a Gaussian instrumental broadening kernel. 

Overall our spectroscopic analysis returned fairly accurate stellar parameters for \emph{Gaia} benchmark standards, and our reduction method provided similar results to the GALAH reduction pipeline (for details see Appendix~\ref{sec:benchmark}). We found no significant offset in our effective temperature or surface gravity compared to reference values derived by \cite{Jofre2014} and \cite{Heiter2015}. The temperature offset is 40 K with standard deviation of 90 K; the surface gravity offset is 0.02 dex with standard deviation of 0.25 dex. The metallicity, however, shows an offset of $-0.12$ with standard deviation 0.08 dex. The metallicity offset is the same as that reported by the GALAH survey~\citep{Sharma2018}. To remain consistent with GALAH, we have added $+0.1$ dex to all of our metallicity values. The standard deviation of the difference between our results and that of benchmark stars can be taken as typical uncertainties in the parameters $T_\mathrm{eff}$ (90 K), $\log g$ (0.25 dex) and [Fe/H] (0.08 dex).  

\begin{figure}
	\centering
	\includegraphics[width=1\columnwidth]{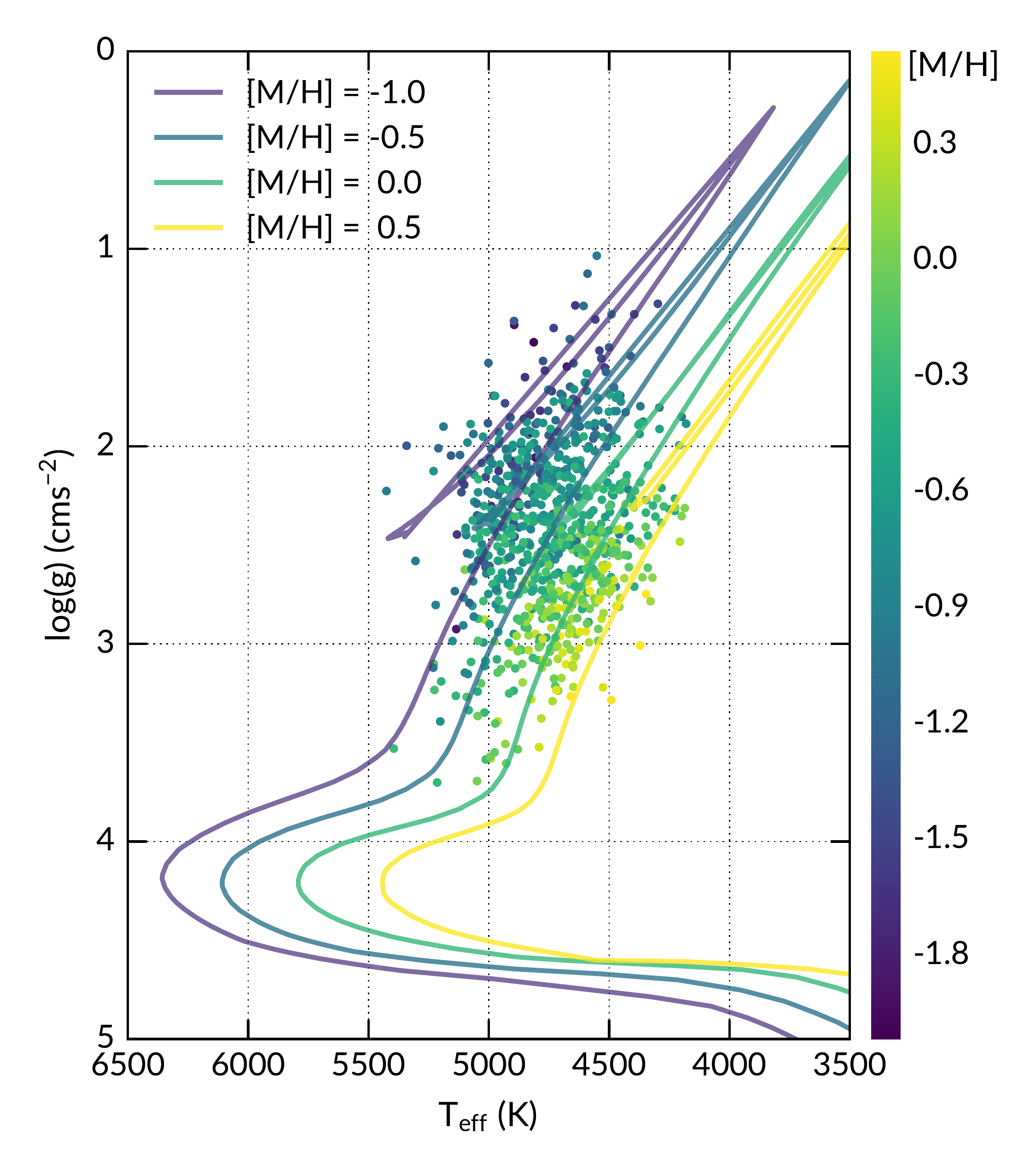}
	\caption{The \emph{Kiel} diagram of the full sample (832 stars), over-plotted with 10 Gyr \textsc{parsec} isochrones~\citep{Marigo2017} of metallicities indicated in the figure legend.}
	\label{fig:HR}
\end{figure}

Fig~\ref{fig:HR} shows the \emph{Kiel} diagram for all minor axis fields. The stellar parameters are well represented by 10 Gyr isochrone tracks, which is what one expects for bulge giants. Of the targets observed ($\approx$350 per field), minus possible binaries and those with reduction issues, we have 313 stars analysed for field $(0,-10)$, 13 of which are from the EMBLA survey. For the pilot field $(0,-7.5)$ there are 315 stars in total. Part of the $(0,-5)$ field was unfortunately affected by very strong hydrogen emission (at H$\alpha$ and H$\beta$ rest wavelengths) from the ISM. This affected the hydrogen line profiles for many stars, and as a result a large fraction of them failed to converge. Therefore, we only have 204 stars in field $(0,-5)$ (two are EMBLA stars), giving us a grand total of 832 stars.

\subsection{Elemental abundances}

After the stellar parameters have been established, they are fixed for each abundance optimisation. Similarly to the parameters, SME optimises the $\chi^2$ parameter to find the best-fit abundance of each element. Only wavelength pixels within the line masks are used for $\chi^2$-minimisation. During the optimisation stage, SME de-selects blended wavelength points (if any) within the line masks. In this paper, we will report on abundances of the alpha elements O, Mg, Si, Ca and Ti. For all elements except Ti, we computed the line-by-line abundances for each element and averaged the individual lines' results, weighted by the abundance ratio uncertainties provided by SME. We do not include in the weighted average abundances that are flagged as upper limits. The Ti abundances are computed with the same lines also used for stellar parameters determination, and all lines were fitted at the same time. In addition, non-LTE corrections were applied to the elements O~\citep{Amarsi2016}, Mg~\citep{Osorio2016} and Si~\citep{Amarsi2017}. 

For solar normalisation, we used abundances from a HERMES twilight spectrum, which was reduced as per Section \ref{redux} and analysed in the same manner as a typical star. This ensures that systematic errors (such as uncertain $\log (gf)$ values) are mostly removed. The solar parameters we derived and adopted for abundance syntheses are: $T_\mathrm{eff}$ = 5735 K, $\log g$ = 4.3 dex, [Fe/H] = $-0.02$ dex, $v_\mathrm{mic}$ = 1.1 km s$^{-1}$. These parameters are different to the nominal solar values from \cite{Prsa2016}, but they are consistent within our estimated uncertainties. We normalised the single-line abundance ratios before computing the weighted average values for each element, such that: 

\begin{equation}
[\mathrm{X/Fe}] = [\mathrm{X/Fe}]_* - [\mathrm{X/Fe}]_{\odot, \mathrm{HERMES}}
\end{equation}

The alpha abundances presented here are not particularly sensitive to temperature, as shown in Fig. \ref{fig:abundteff}. We do not see any appreciable trends with $T_\mathrm{eff}$ for the elements O, Si, and Ti. However, linear trends can be seen for Mg and Ca, which are also observed in GALAH data~\citep{Buder2018}. While we note these issues, we do not apply empirical corrections to abundance-temperature trends, as the underlying physics is yet to be understood, and should be investigated further.

\section{ARGOS comparison and metallicity distribution functions}
\label{sec:argoscomp}

\begin{figure}
	\centering
	\includegraphics[width=0.95\columnwidth]{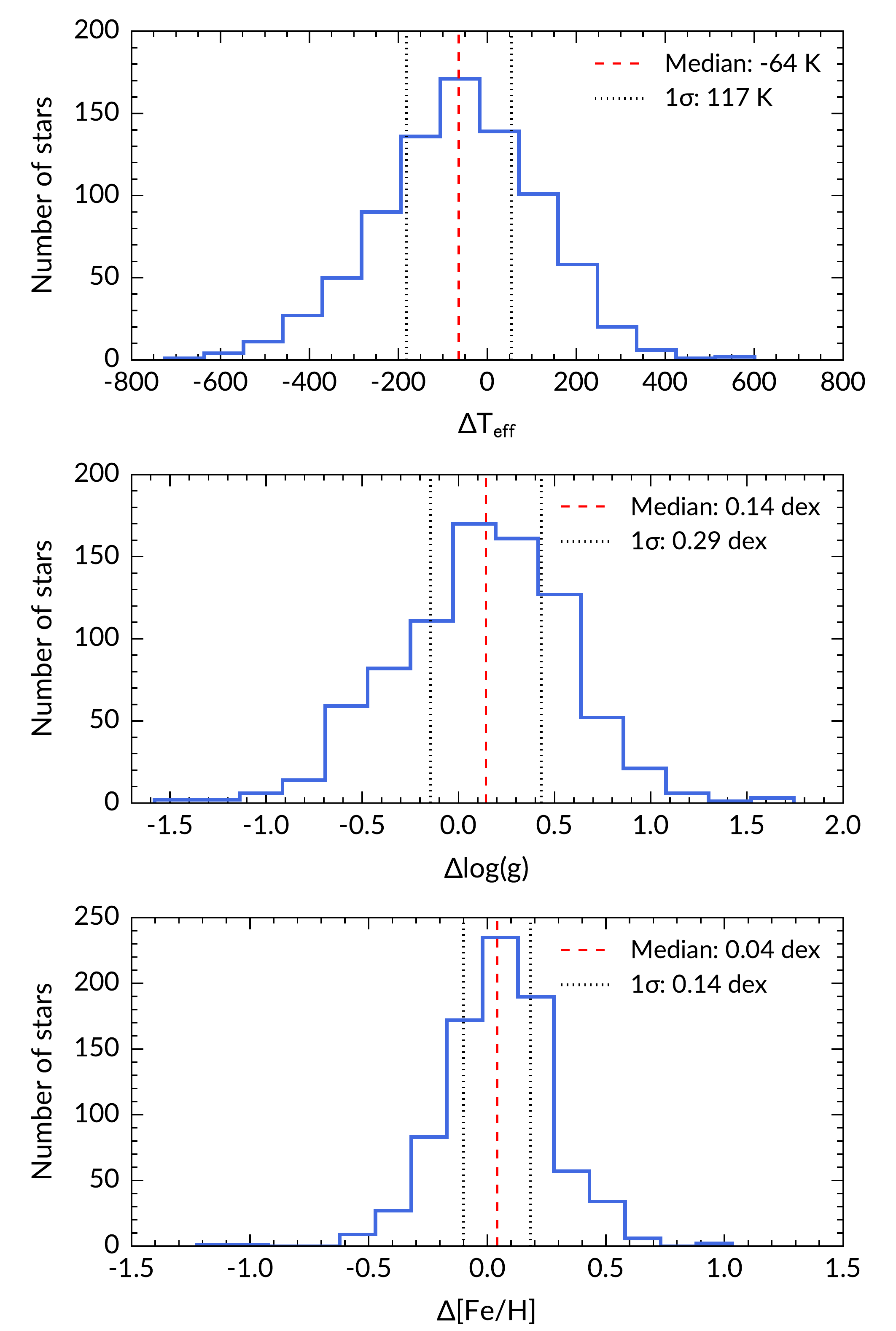}
	\caption{Histogram comparison between parameters derived in this work and those of ARGOS for stars in common. The differences are shown as (HERBS $-$ ARGOS). The 1$\sigma$ levels (dotted lines) are median absolute deviations (MAD), which are more robust against outliers.}
	\label{fig:argoscomp}
\end{figure}

\begin{figure}
	\centering
	\includegraphics[width=1\columnwidth]{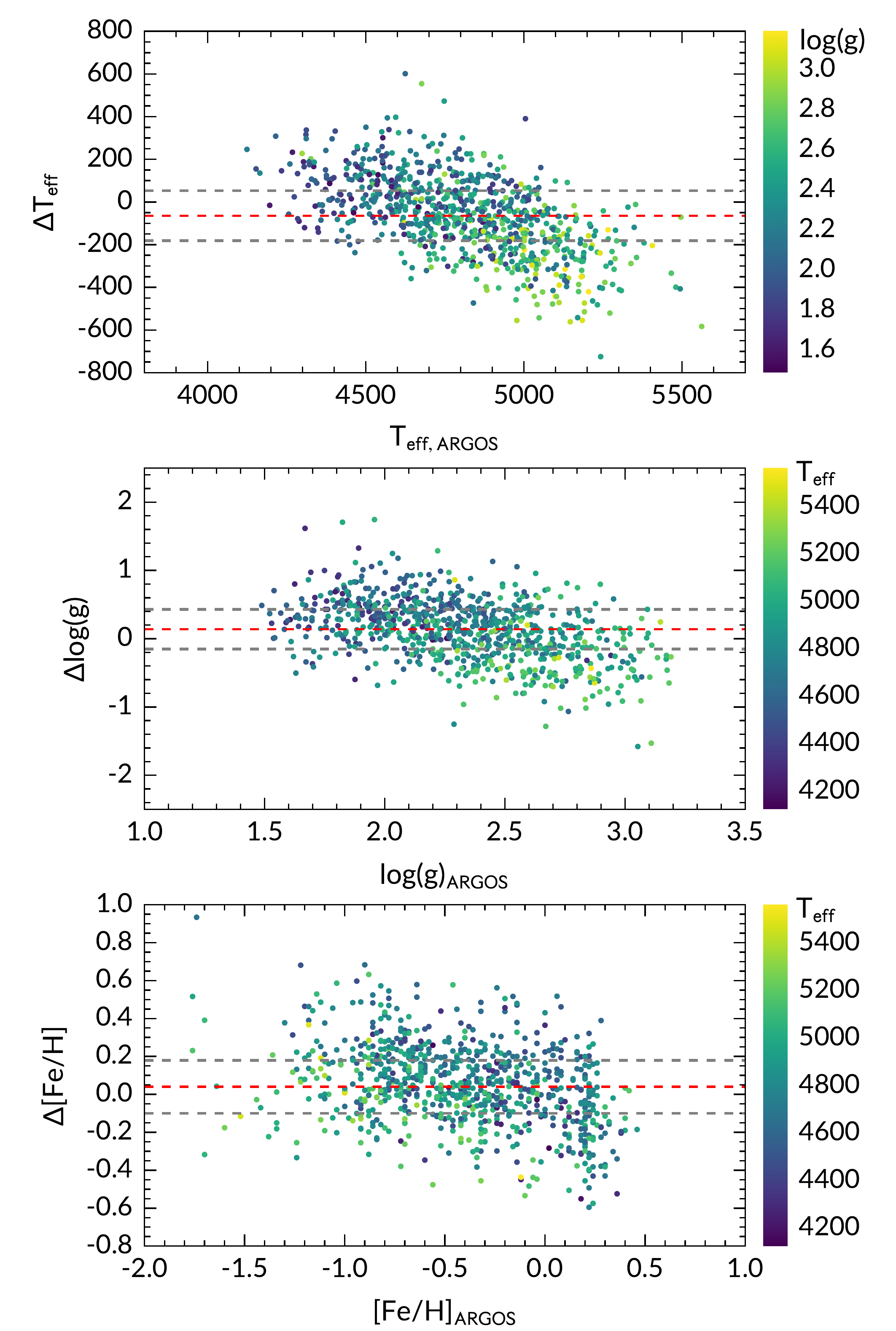}
	\caption{The same (HERBS $-$ ARGOS) differences as Fig. \ref{fig:argoscomp}, but plotted as a function of ARGOS stellar parameters, colour-coded by $T_\mathrm{eff}$ or $\log g$. Lighter colours represent warmer stars with higher surface gravities.}
	\label{fig:argoscomp2}
\end{figure}

As noted earlier, we selected most of our bulge stars from the ARGOS survey, which was observed with the AAOmega spectrograph ($\lambda/\Delta \lambda = $11 000). Fig. \ref{fig:argoscomp} and \ref{fig:argoscomp2} show the comparison between our parameters and that of ARGOS for stars in common. In Fig \ref{fig:argoscomp}, the differences are plotted as histograms: the biases (median of the difference) are shown for $T_\mathrm{eff}$, $\log g$ and [Fe/H]. {The 1$\sigma$ values in the figures were computed using the median absolute deviation (MAD) method, which is more robust to outliers. The standard deviations, after excluding 3$\sigma$ outliers are: $\Delta T_\mathrm{eff}$: 158 K, $\Delta \log g$: 0.38 dex and $\Delta$[Fe/H]: 0.16 dex.} This comparison shows small offsets between the two studies. The standard deviation in $\Delta$[Fe/H] is consistent with the combined HERBS and ARGOS metallicity uncertainties, and the overall bias is negligible. Even though ARGOS effective temperatures were determined using photometry ($J-K_0$ colours), they agree remarkably well with our values {and the standard deviation of $\Delta T_\mathrm{eff}$ is in line with the two studies' combined uncertainty. Because the ARGOS photometric temperatures are less accurate at low latitudes due to increased extinction, field $(0,-5)$ has a higher MAD value (143 K) compared to fields $(0,-7.5)$ and $(0,-10)$ (106 K and 111 K, respectively). Surface gravity shows a small 0.14 dex offset, but the standard deviation is expected of the combined HERBS and ARGOS $\log g$ uncertainties.}

\begin{figure*}
	\centering
	\includegraphics[width=1\columnwidth]{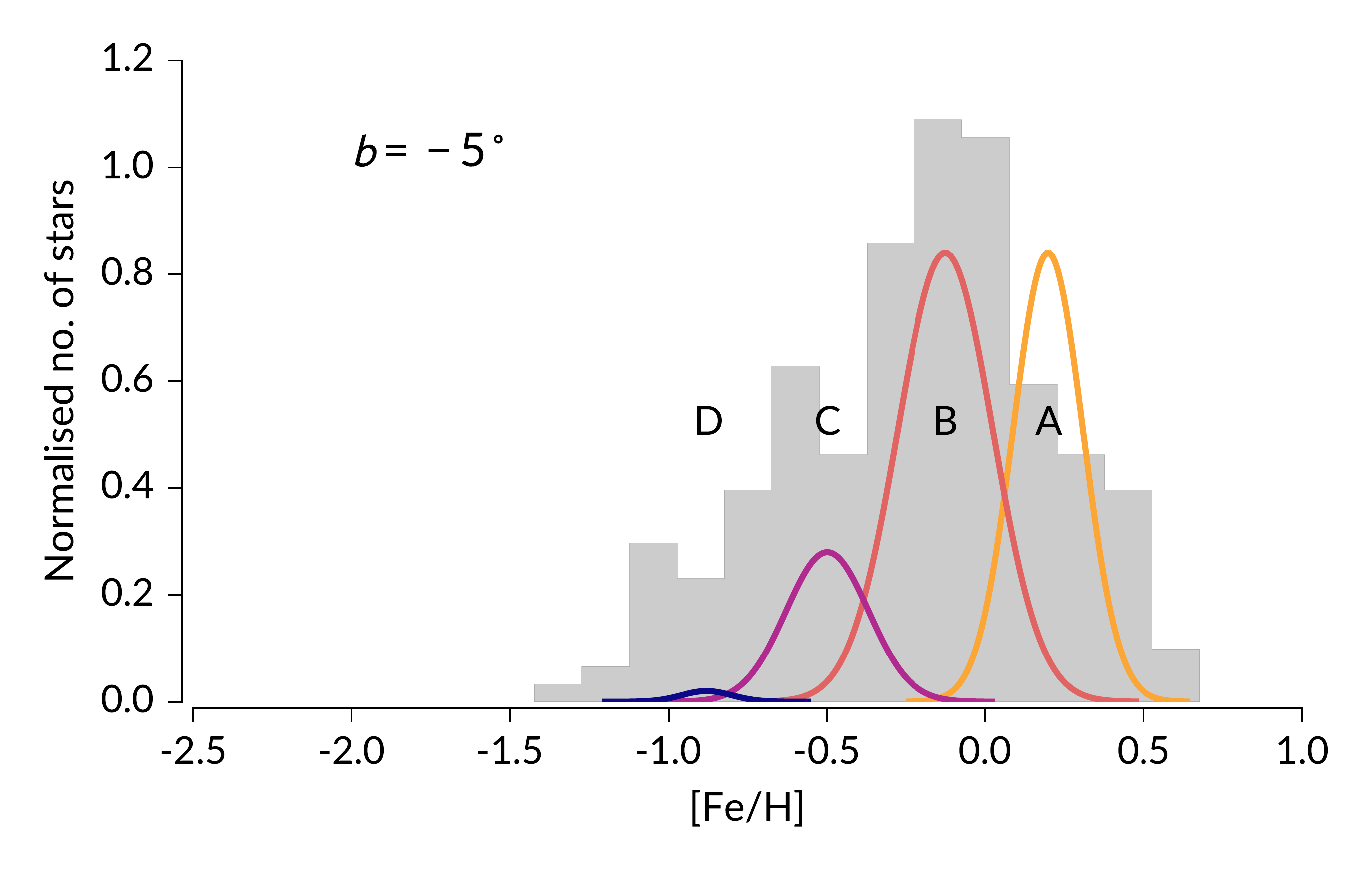}\\
	\includegraphics[width=1\columnwidth]{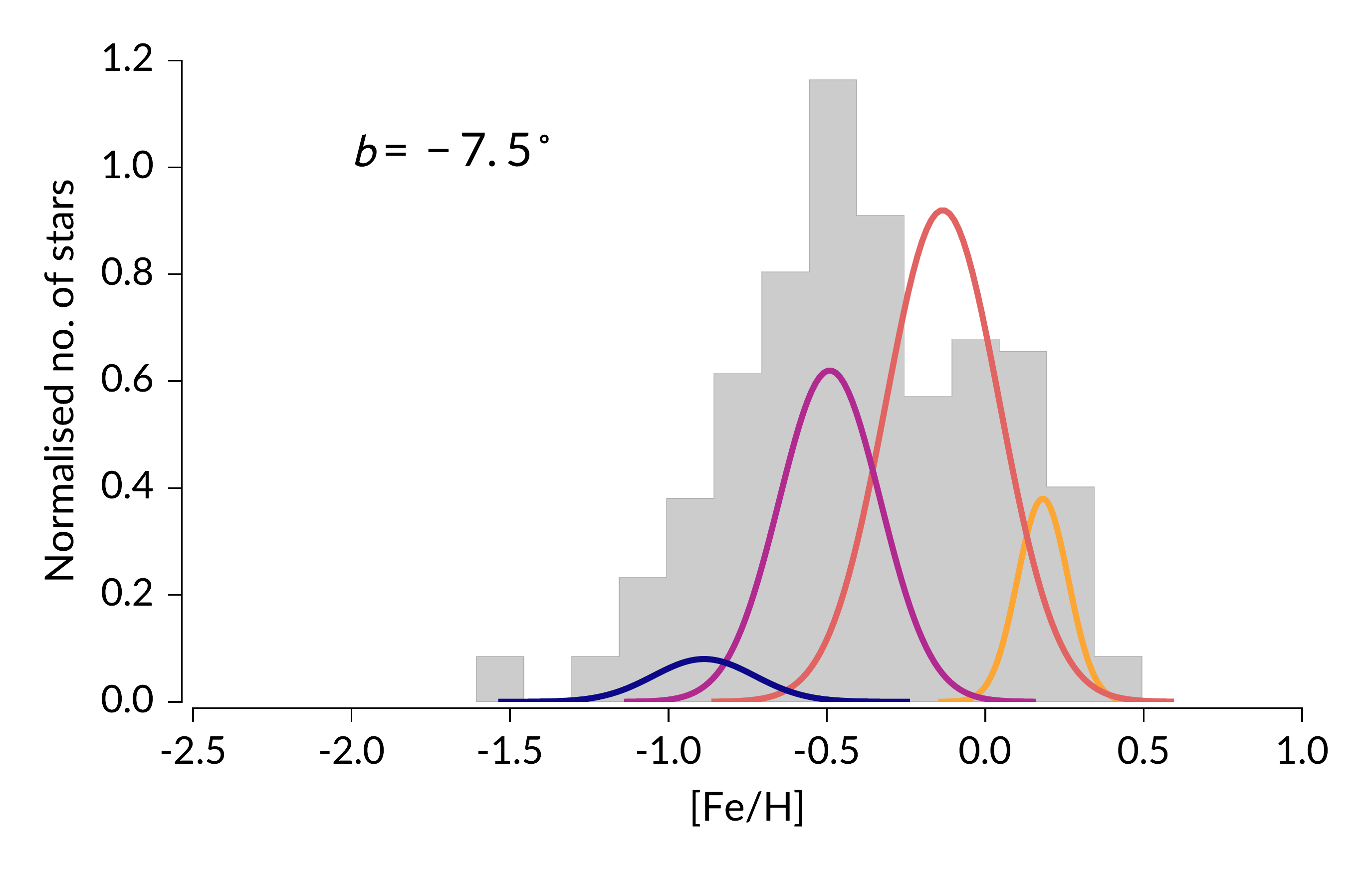}\includegraphics[width=1\columnwidth]{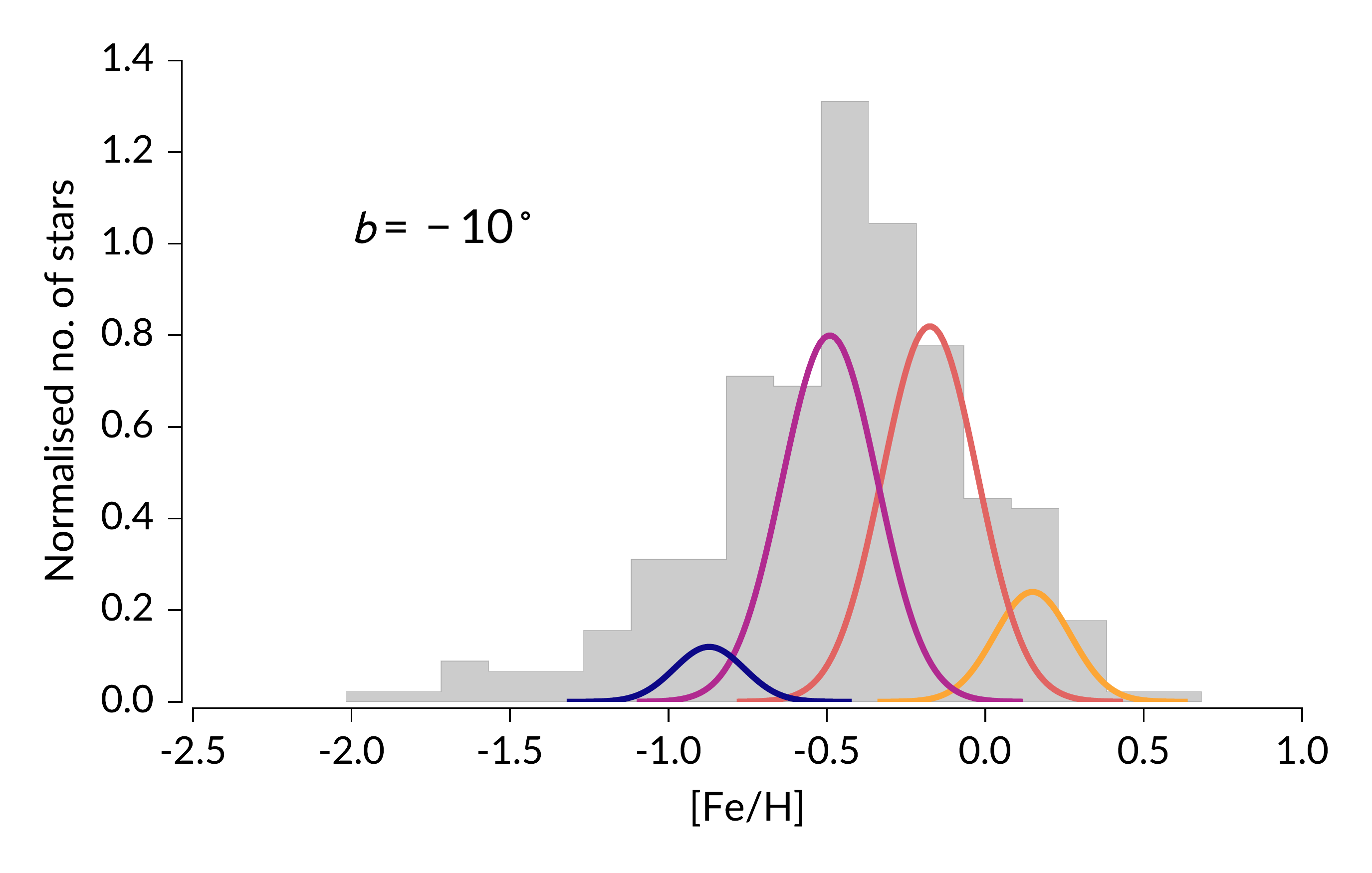}
	\caption{The metallicity distribution functions (grey histograms) for the minor axis fields. Over-plotted for comparison are ARGOS metallicity components A--D (corresponding to most metal-rich to most metal-poor). The ARGOS centroids have been shifted according to Equation \ref{diffeqn}. The amplitude of each component has been multiplied by a factor of two to better match the distributions presented here, but their relative weights remain the same.}
	\label{fig:mdf}
\end{figure*}

Fig \ref{fig:argoscomp2} shows the same differences as Fig. \ref{fig:argoscomp}, but as functions of ARGOS stellar parameters. In this figure, trends as a function of parameters are apparent. The trend in temperature could have been caused by the photometric calibration that was used to determine ARGOS effective temperatures. In general, stars with higher ARGOS $T_\mathrm{eff}$ and $\log g$ are estimated to be cooler, and have lower surface gravity in our analysis. For [Fe/H], there is a mild linear trend, which is not apparently dependent on $T_\mathrm{eff}$ or $\log g$. The trend in $\Delta$[Fe/H] can be described as:
\begin{equation}
\label{diffeqn}
\Delta[\mathrm{Fe/H}] = -0.190(0.013) \times [\mathrm{Fe/H}]_\mathrm{ARGOS} - 0.04(0.01)
\end{equation}

\noindent In Equation~\ref{diffeqn}, the numbers in parentheses indicate the uncertainties of the slope and intercept. The trend indicates that metal-rich ARGOS stars are estimated to be slightly more metal-poor in our analysis, and vice versa.

The metallicity distribution for each minor-axis field is shown in Fig. \ref{fig:mdf}. Here we have also over-plotted the Gaussian distributions corresponding to each ARGOS component A--D; A being the most metal-rich, D the most metal-poor. As indicated above, our analysis suggests a slight compression of the ARGOS MDF. In Fig. \ref{fig:mdf}, we have shifted the centroids of these components according to Equation \ref{diffeqn} to reflect this compression. 

On the whole, our MDF appears flatter and wider compared to the ARGOS MDF, however this is to be expected, as our selection function prioritised the most metal-rich and metal-poor stars. The selection criteria we employed have allowed for a larger fraction of metal-poor star to be observed. Approximately 12\% of the stars have [Fe/H] $\leq$ -1, compared to the typical fraction of 4--5\%~\citep{Ness2013, Rojas-Arriagada2017}. 

There are discussions in the literature regarding the number of metallicity components in the bulge, with some authors arguing for a two-component bulge metallicity distribution with much larger dispersions~\citep{Gonzalez2015a, Rojas-Arriagada2017, Schultheis2017}, rather than three components with narrow dispersions~\citep{Ness2013,Perez2018}. As there are strong selection effects associated with our MDF, we are not able to directly address this issue. However, our analysis indicates that the ARGOS component centroids should be located $\approx$0.09 dex closer together. This difference is sufficiently small that for fields $(0,-5)$ and $(0,-10)$, the ARGOS components remain distinct given their narrow dispersions. However, for field $(0,-7.5)$, the centroid of component A is within $1.5\sigma$ of component B's centroid, meaning that there is a possibility components A and B are not distinct in field $b=-7.5^\circ$.

It is worth noting that the number of metallicity components may not be indicative of how many distinct populations reside in the Galactic bulge. Indeed, the N-body dynamical model of~\cite{Fragkoudi2018} found that even though their bulge population originated from three different disk components, the final MDF is best described by two Gaussian curves with larger dispersions than the original disk components.

\section{$\alpha$-element abundances} 

\begin{figure*}
	\centering
	\includegraphics[width=1.3\columnwidth]{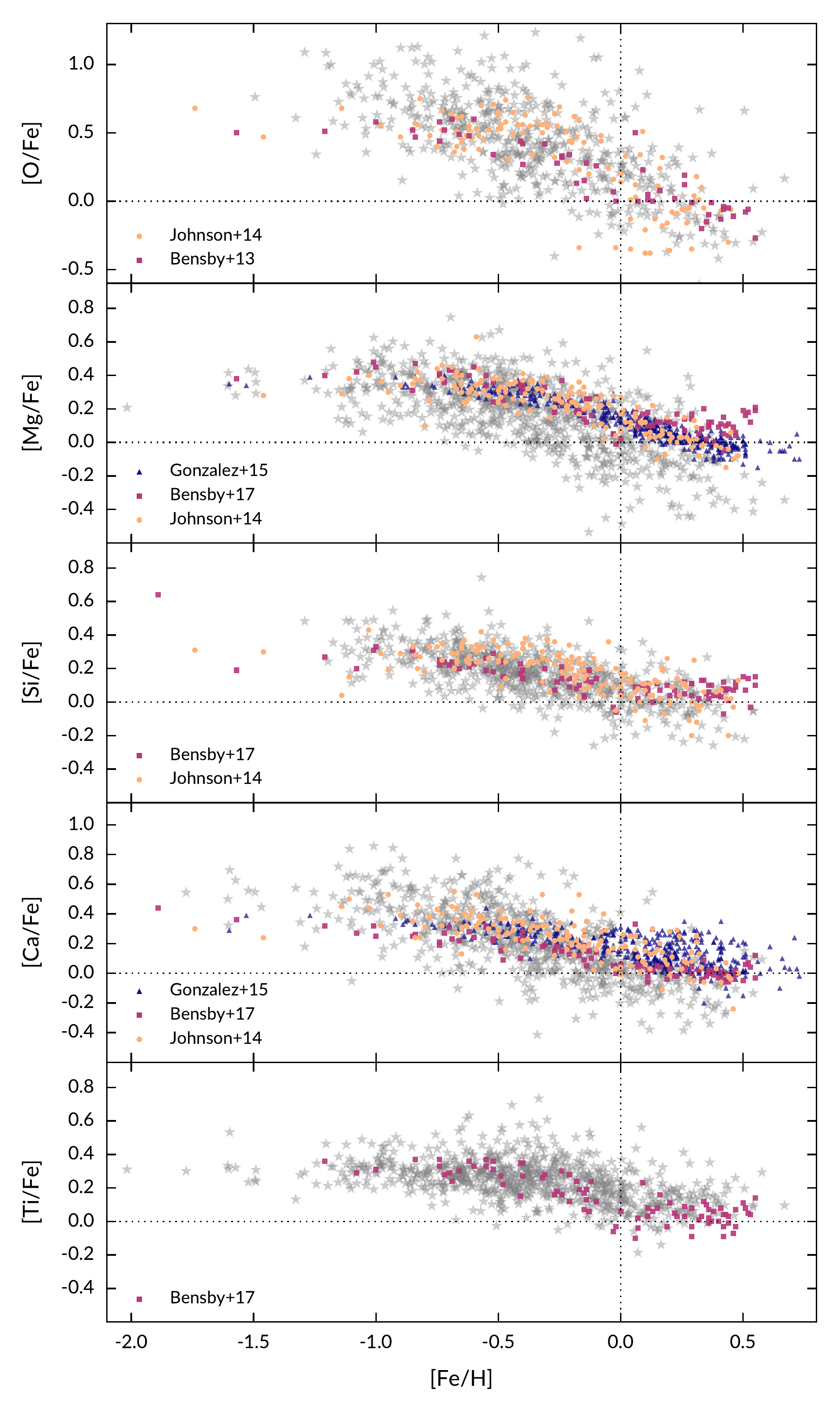}\\
	\caption{Abundance ratios for the alpha elements from this work (grey stars) compared to literature studies. The abundance trends derived here are largely in agreement with the literature. In particular, Si, Ca and Ti seem to follow the same trend and similar abundance scale to the microlensed dwarfs~\protect\citep{Bensby2017}. However, for the elements O and Mg, our scatter is considerably larger than other studies. See text for details.}
	\label{fig:litcomp}
\end{figure*}

The alpha elements are often associated with rapid SNeII enrichment, and are thus useful indicators of formation/evolution timescales for the different Galactic components~\citep{Tinsley1979,Matteucci1990}. Fig \ref{fig:litcomp} shows the alpha abundances from this study compared to recent high-resolution spectroscopic studies of bulge \emph{field} stars in the literature. For this exercise we have included \cite{Bensby2013,Bensby2017}, \cite{Johnson2014} and \cite{Gonzalez2015a}. The \cite{Bensby2017} study includes all microlensed bulge dwarfs in \cite{Bensby2013}, however we use oxygen abundance ratios from \cite{Bensby2013}, which is not available in the later study. The microlensed dwarfs were observed with the VLT/UVES spectrograph; KECK/HIRES spectrograph or Magellan/MIKE spectrograph ($\mathcal{R} \approx$ 40 000--90 000). Both \cite{Johnson2014} and \cite{Gonzalez2015a} provided individual abundances for a large number of bulge giants observed with the VLT/GIRAFFE spectrograph ($\mathcal{R} \approx$ 22 500), but at different wavelength settings. All literature samples considered are smaller, but have on average higher SNR than our sample. Furthermore, only the oxygen abundance ratios from \cite{Bensby2013} were computed assuming non-LTE. All studies assumed LTE in their abundance analysis. On the whole, the scatter in HERBS abundance ratios are larger than the comparison samples and what could be expected from abundance ratio uncertainties from $\chi^2$-square fitting. This indicates that the $\chi^2$-square errors may be underestimated. We describe the trends of each element below.
\begin{itemize}
\item \textbf{Oxygen} The oxygen abundance trend is largely in agreement with literature studies, but with much larger scatter. We used the \ion{O}{i} line at 7772 \AA, which is the strongest line of the triplet used by \cite{Bensby2013}\footnote{It made little difference to [O/Fe] whether we use all three lines of the oxygen triplet, or just the 7772 \AA~line. As the other two lines of the triplet are significantly weaker (therefore often undetectable) in our spectra, we chose to use only the 7772 \AA~line.}. Given that the line strength of the oxygen triplet is weaker in giants than in dwarfs, and the lower SNR of our spectra, it is not surprising that our scatter is larger than that of \cite{Bensby2013}. In addition, the plateau of [O/Fe] is not as well defined as in other works, but [O/Fe] decreases as a function of metallicity, from [Fe/H] $\approx -0.4$ dex. As other authors have commented, the average oxygen abundance ratio is higher, and the decline of [O/Fe] with metallicity is steeper than that of other alpha elements~\citep{Bensby2013, Johnson2014}. This indicates that aside from SNeIa contribution of iron, other mechanisms may have affected the decrease in [O/Fe], such as stellar mass loss~\citep{McWilliam2008,McWilliam2016}. 

\item \textbf{Magnesium} Similarly to oxygen, magnesium largely follows the same trend as literature studies, but with larger scatter. This could largely be attributed to the low SNR of our spectra. It is also apparent that the mean [Mg/Fe] of this study is lower than that of other studies by $\approx$0.15 dex. Although \cite{Johnson2014}, \cite{Gonzalez2015a} and \cite{Bensby2017} analysed different types of stars, using different methods, their abundance trends and scale agree well up to solar metallicity. At [Fe/H] $\approx$ 0, \cite{Bensby2017} showed a flattening trend for [Mg/Fe], while \cite{Johnson2014} and \cite{Gonzalez2015a} showed continued decrease as a function of [Fe/H]. Our results are in line with the latter trend. 

\item \textbf{Silicon} For Si, our abundance trend seems to follow that of \cite{Bensby2017}, but with slightly larger scatter. \cite{Johnson2014} estimated on average higher [Si/Fe] compared to this work and \cite{Bensby2017} at [Fe/H] $ < 0$ dex. All three studies are in agreement that [Si/Fe] flattens at super-solar metallicity to approximately the solar value, however \cite{Bensby2017} observe slightly more enhanced [Si/Fe] in this regime. 

\item \textbf{Calcium} For Ca, the general abundance trend is consistent with all three literature samples. However, at sub-solar metallicity, our [Ca/Fe] values are in agreement with \cite{Johnson2014} and \cite{Gonzalez2015a}, which are on the mean higher than those reported by \cite{Bensby2017}. Both \cite{Johnson2014} and \cite{Gonzalez2015a} find enhanced [Ca/Fe] at super-solar metallicity, but \cite{Gonzalez2015a} reported rather large uncertainties for [Ca/Fe] in this regime. Here, our results seem to be in good agreement with \cite{Bensby2017}, with [Ca/Fe] flattening to solar value for [Fe/H] $ > 0$. {However, there is a small offset (0.05 dex, ours being lower) between our results and that of \cite{Bensby2017}. The scatter in our [Ca/Fe] measurements is $\approx$0.2 dex, somewhat higher than other studies.} 

\item \textbf{Titanium} The [Ti/Fe] trend derived here is in good agreement with \cite{Bensby2017}, but different from the trends established for giants by e.g.,~\cite{Alves-Brito2010} and \cite{Gonzalez2011}. Both the microlensed dwarfs and our giants show that [Ti/Fe] decreases to near-solar value at [Fe/H] $\approx$ 0 and flattens at super-solar metallicity, consistent with the behaviours of silicon and calcium. However, our [Ti/Fe] values remain enhanced by $\lesssim$0.1 dex compared to \cite{Bensby2017} at super-solar metallicity.

\end{itemize}

In summary, the alpha abundances derived in this work follow the same trend as some of the most recent, high signal-to-noise, high-resolution studies of bulge stars. For oxygen and magnesium, our results show considerably larger scatter (approximately twice) compared to literature studies, and an offset in the mean magnesium abundances. However, [Si/Fe], [Ca/Fe] and [Ti/Fe] show comparable scatter and abundance scale to other studies. For both O and Mg, a plateau can be seen from [Fe/H] $\lesssim -0.5$, and the abundances of both elements decrease as function of [Fe/H] above solar metallicity. Si, Ca and Ti show similar trends, flattening to near-solar or solar values for [Fe/H] $\geq 0$. This behaviour at super-solar metallicity was seen as unique to microlensed bulge dwarfs (e.g., \citealt{Johnson2014}), however we confirm that this is not the case. A plateau at [X/Fe] $\approx 0.3$ dex can be seen for Si and Ti, from [Fe/H] $\lesssim -0.5$. Finally, although the abundance trends are largely consistent with the literature, there are some inconsistencies in terms of abundance scale at different metallicity regimes. 

\subsection{Variation with latitude}

\begin{figure*}
	\centering
	\includegraphics[width=1.05\columnwidth]{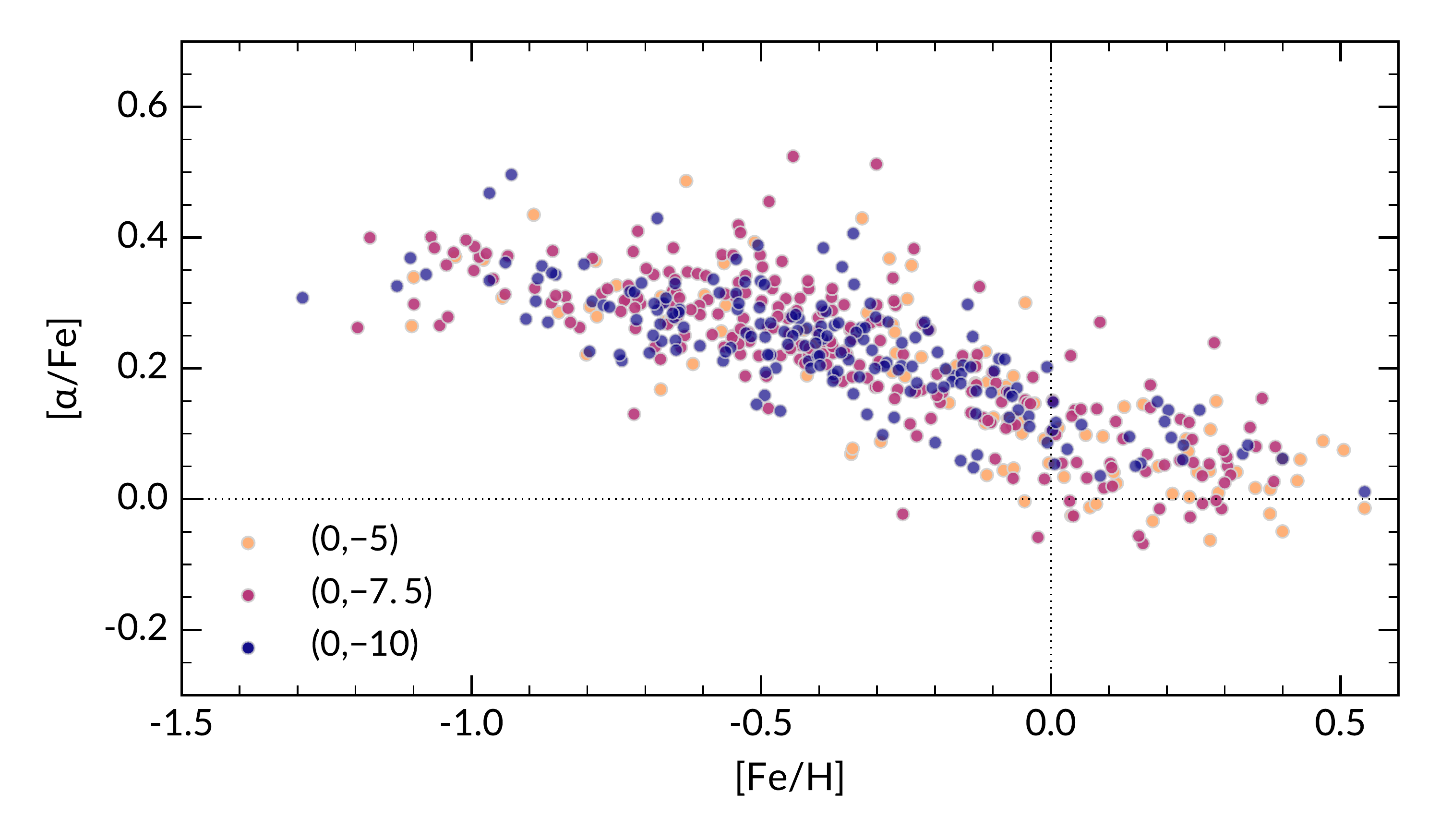}\includegraphics[width=1.05\columnwidth]{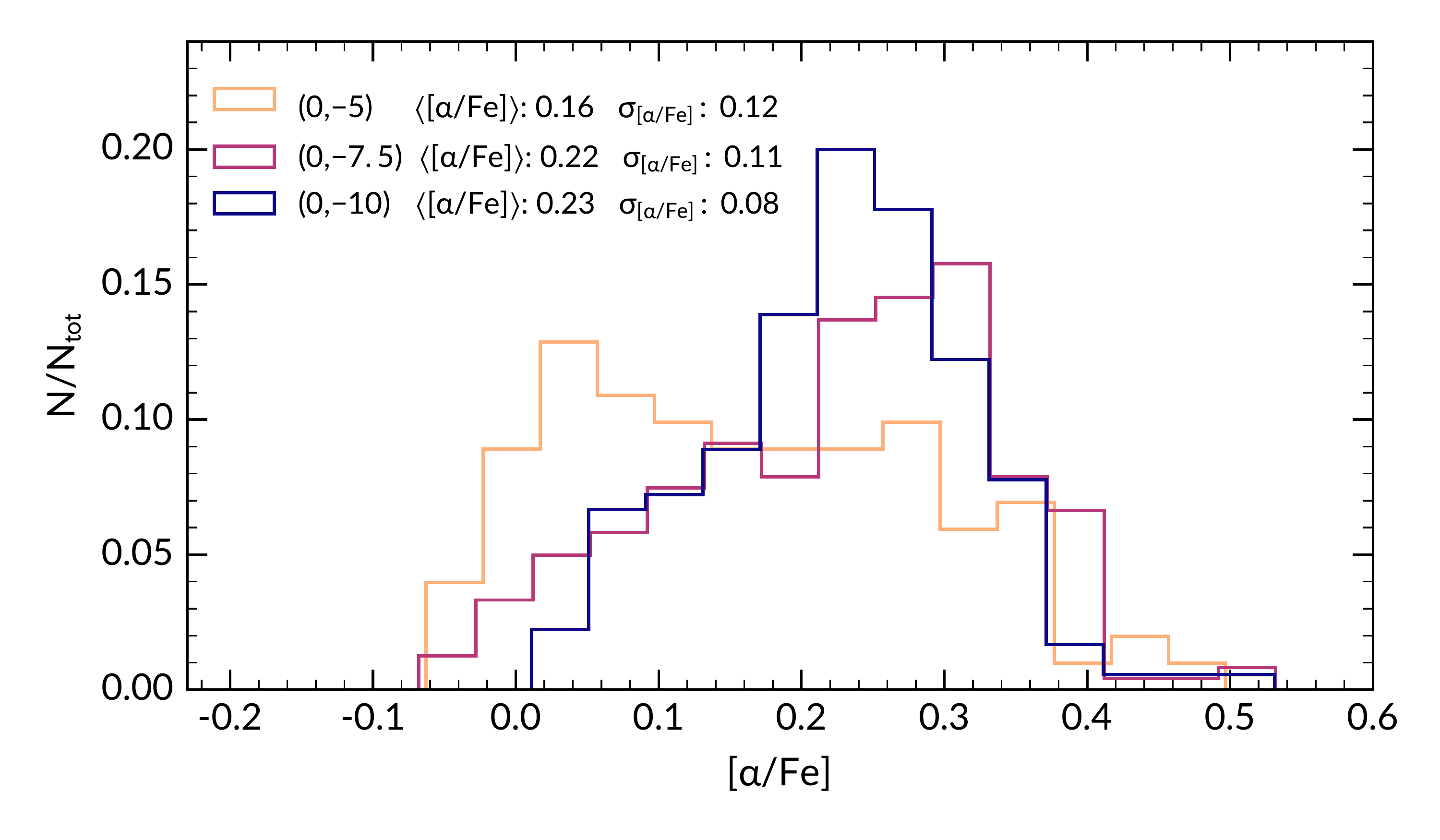}
	\caption{Left panel: the weighted average of Mg, Si, Ca and Ti ([$\alpha$/Fe]) as a function of [Fe/H] for the three minor axis fields. Right panel: the corresponding histograms of [$\alpha$/Fe] at different latitudes. The mean [$\alpha$/Fe] values and standard deviations are given for each field. Overall [$\alpha$/Fe] increases with latitude, but the dispersion seems to be smaller at the highest latitude ($b=-10$).}
	\label{fig:avgalpha}
\end{figure*}

As the contribution of metallicity components change with latitude, such that the metal-rich component dominates near the plane, one would expect a vertical gradient in [$\alpha$/Fe] in the opposite sense: that the low-$\alpha$ population dominates near the plane, and the high-$\alpha$ population dominates away from the plane. This has been observed by \cite{Gonzalez2011} for their bulge giants located near the minor axis, at Baade's window and $(\ell,b)=(0.21,-6) $ and $(0,-12)$. More recently, \cite{Fragkoudi2018} showed that a positive [$\alpha$/Fe] gradient is also present in their N-body simulation, where the bulge population originated from three disk components. 

The same conclusion can be drawn from Fig. \ref{fig:avgalpha}, which shows the distribution of [$\alpha$/Fe] for bulge fields observed in this work. We used the weighted average of the elements Mg, Si, Ca and Ti to determine [$\alpha$/Fe]. Oxygen was excluded because it may have a different chemical evolution history to the other alpha elements, as discussed in the previous section. The weighted average [$\alpha$/Fe] is mostly influenced by Si and Ti, which are the most precisely measured elements. We do not report [$\alpha$/Fe] values for [Fe/H] $< -1.5$ because most of the elemental abundances cannot be measured at this metallicity regime. 

We observe that the mean [$\alpha$/Fe] indeed shifts towards higher values at higher latitudes. However, the median value and shape of [$\alpha$/Fe] changes sharply between $b=-5^\circ$ and  $b=-7.5^\circ$. Closer to the plane, the distribution is fairly uniform, but away from the plane, it is positively skewed (towards higher [$\alpha$/Fe] values). The alpha abundance distributions of $b=-7.5^\circ$ and $b=-10^\circ$ are similar in shape and mean value, however $b=-7.5^\circ$ have slightly larger dispersion. These observations can be explained by the relative contributions of different metallicity components observed by ARGOS: the fraction of the most metal-rich (low-$\alpha$) component drops significantly between $b=-5^\circ$ and $b=-7.5^\circ$, whereas the component contributions are similar for fields $b=-7.5^\circ$ and $b=-10^\circ$~\citep{Ness2013a}. Qualitatively, our results are consistent with that of \cite{Gonzalez2011} (see their Figure 15), {although our MDF is biased, which could affect the [$\alpha$/Fe] distribution function}. An interesting point to note is that there is a hint of decreasing $\sigma_{[\alpha/\mathrm{Fe}]}$ as a function of latitude, which can also be seen for southern APOGEE bulge fields near the minor axis~(See \citealt{Fragkoudi2018}, Figure 12). 

\begin{figure*}
	\centering
	\includegraphics[width=1\columnwidth]{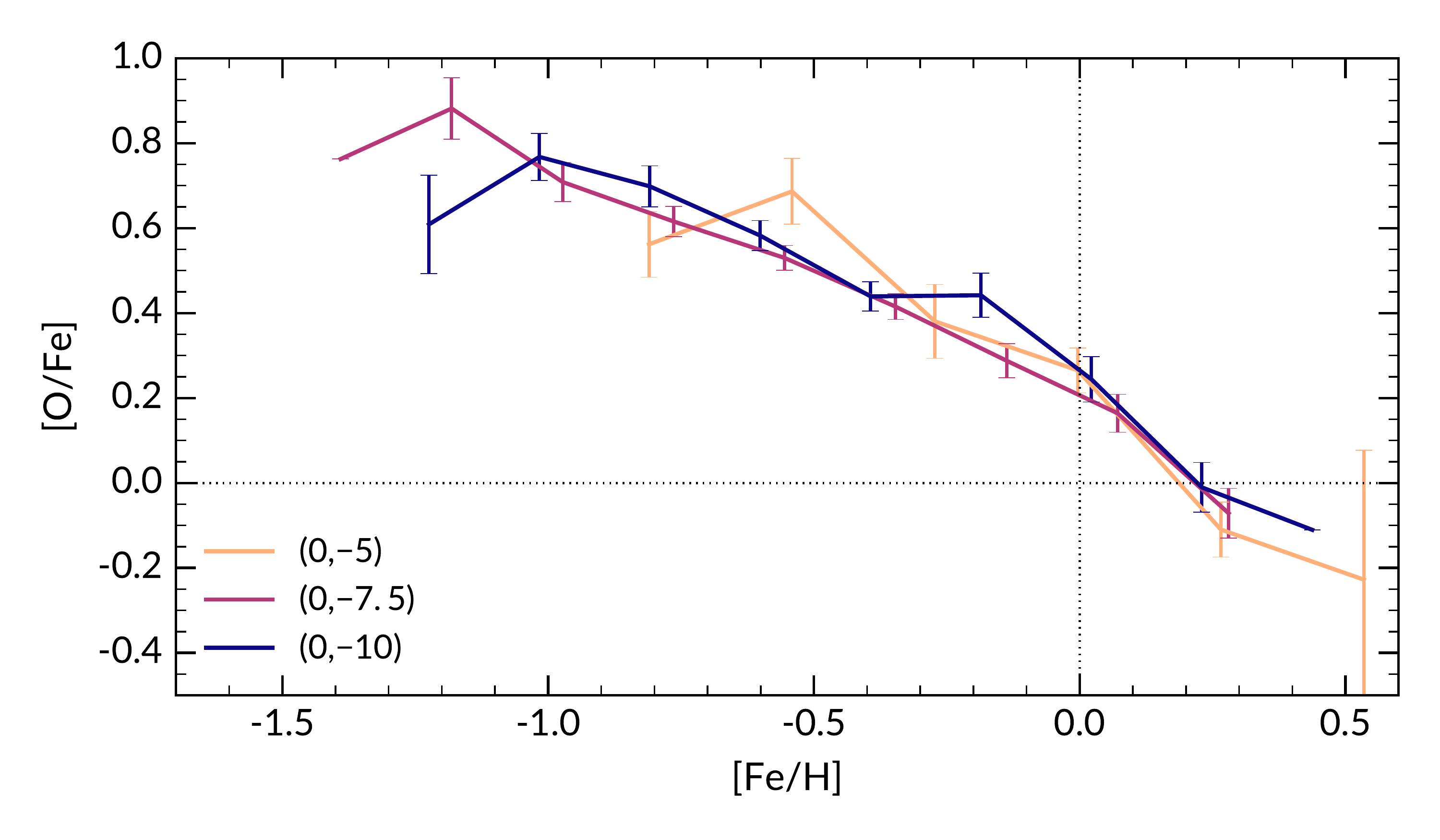}\includegraphics[width=1\columnwidth]{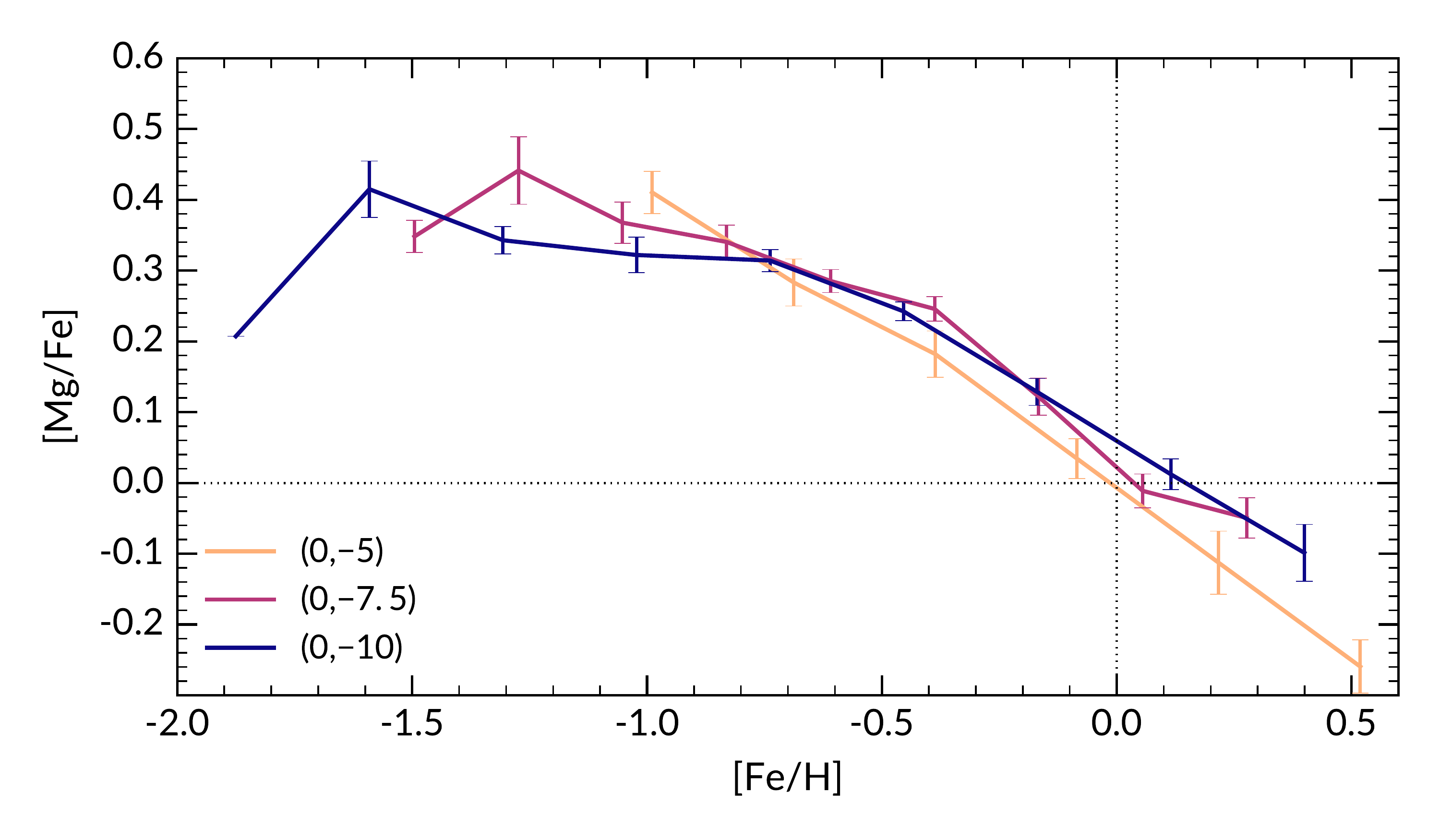}\\
	\includegraphics[width=1\columnwidth]{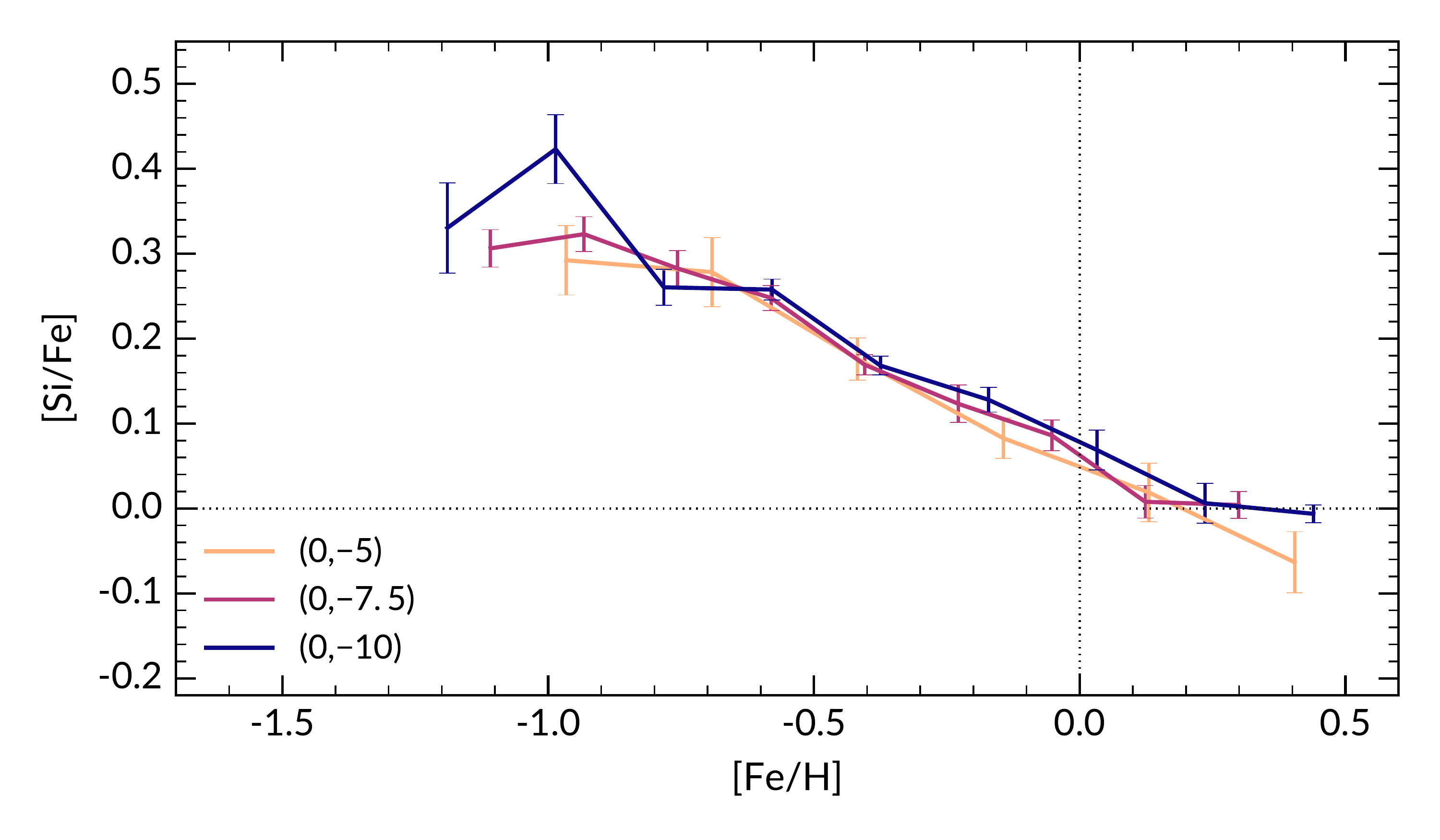}\includegraphics[width=1\columnwidth]{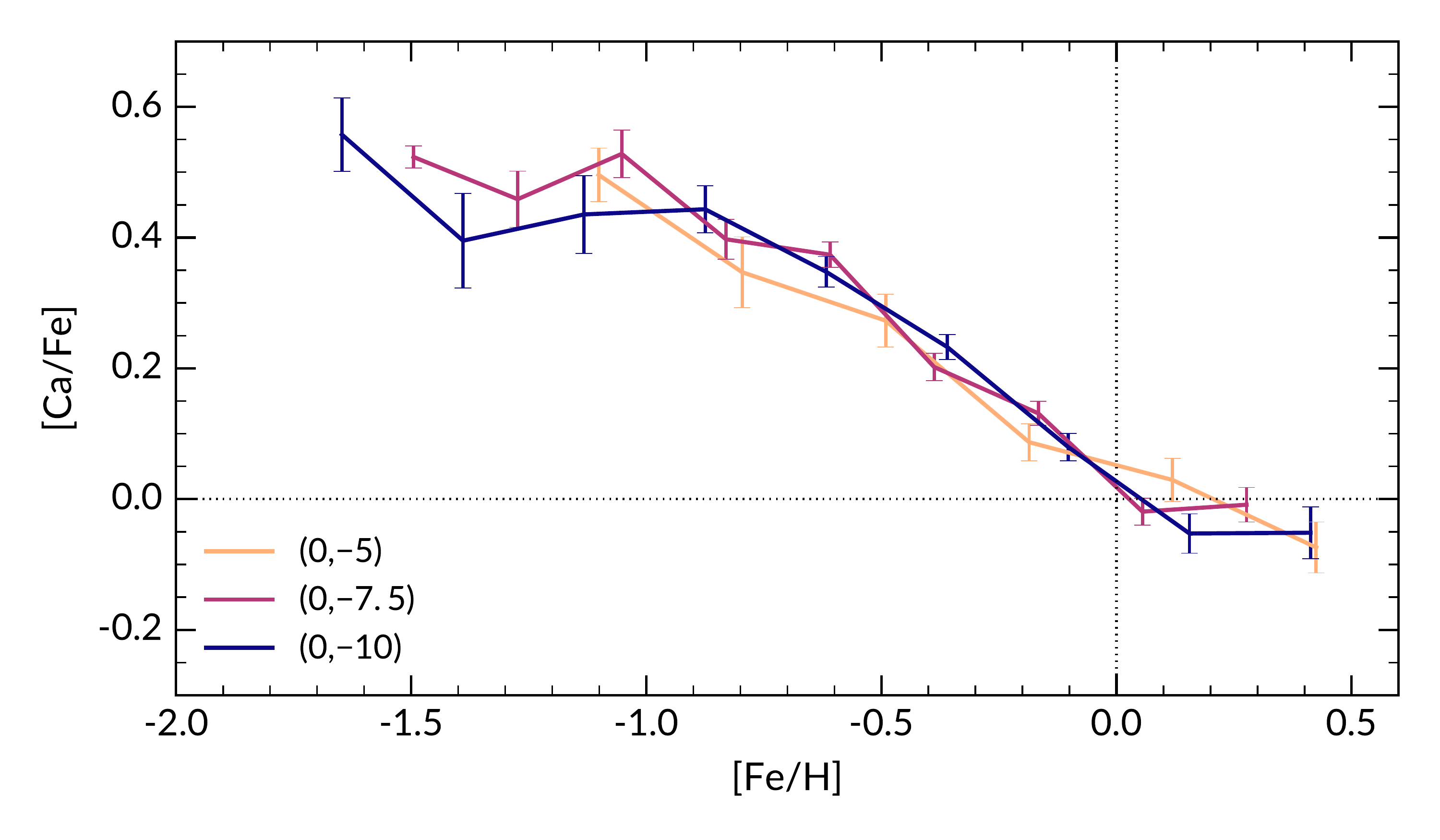}\\
	\includegraphics[width=1\columnwidth]{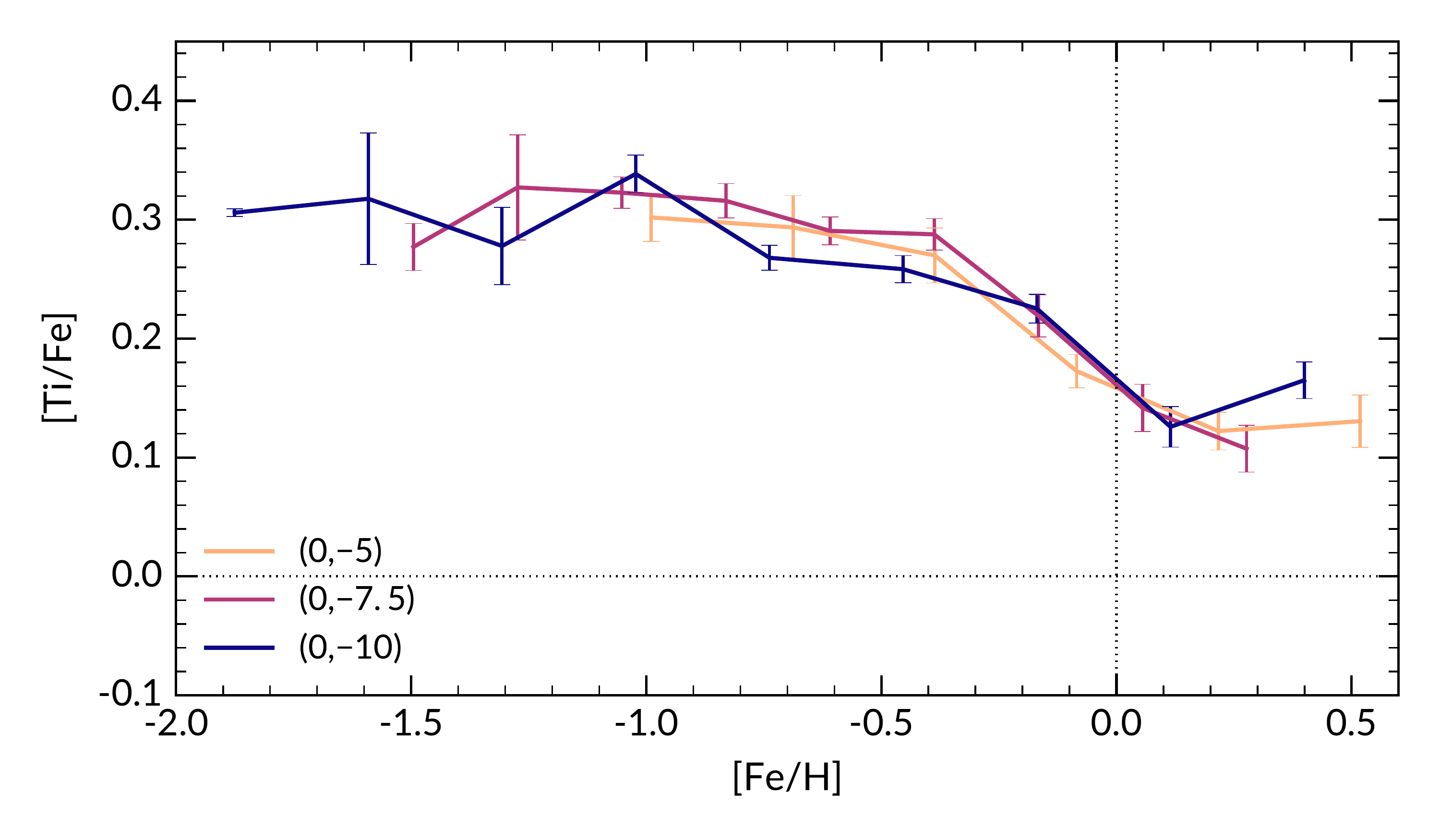}\includegraphics[width=1\columnwidth]{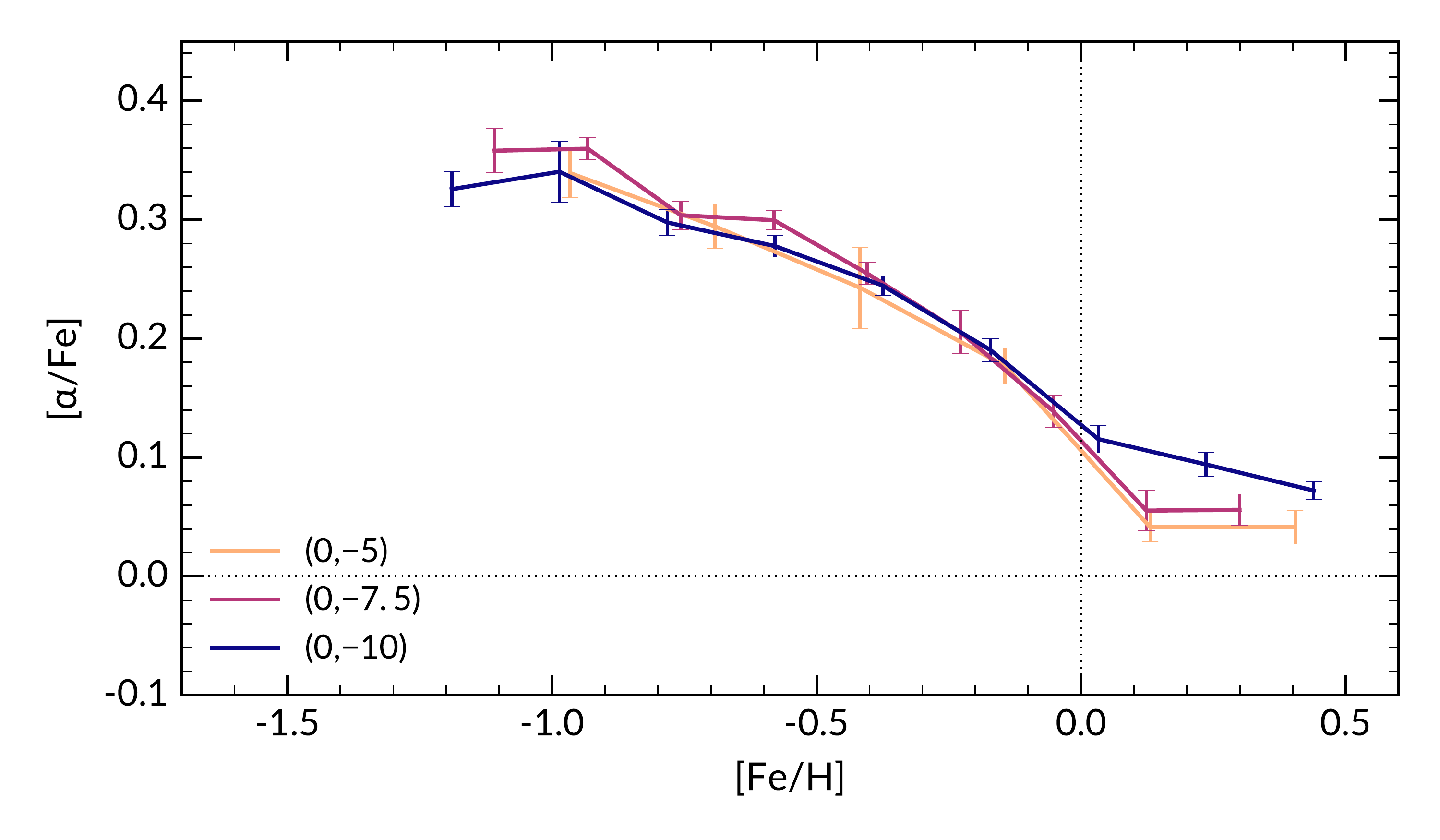}
	\caption{The median trends of each $\alpha$-element, and the weighted average [$\alpha$/Fe] at different latitudes. Median points are computed for [Fe/H] bins of $\approx$0.2 dex in width. The error bars are the standard deviation in the mean.}
	\label{fig:abundlat}
\end{figure*}

While an increase in the mean [$\alpha$/Fe] with distance from the plane is expected~\citep{Gonzalez2011,Fragkoudi2018}, intrinsic vertical abundance gradients of the different bulge metallicity components have not been established. In this work we are well placed to assess the variation with latitude (if any) of each metallicity component, as we have sampled the same range of metallicity at each latitude. For some elements, we are able to measure abundances down to [Fe/H] $\approx -2$, but for most elements (and thus [$\alpha$/Fe]), we are able to probe metallicity components within $ -1.5 \lesssim \mathrm{[Fe/H]} \lesssim 0.5$.

For each latitude, we computed the median [$\alpha$/Fe] at different [Fe/H] bins, in steps of 0.2 dex, shown in Fig. \ref{fig:abundlat}. Errors in the median [$\alpha$/Fe] values are computed as the standard error in the mean. Given the uncertainties, we do not observe variations in alpha abundances across the latitude range covered here for $-1 <$ [Fe/H] $<$ 0. Similarly, \cite{Ryde2016} did not find vertical variations in the alpha abundances of inner bulge stars within two degrees from the Galactic plane. At the low metallicity regime ([Fe/H] $<-1$), variations between fields $(0,-7.5)$ and $(0,-10)$ can be seen for certain elements (O, Mg and Si). However, the trends are not consistent. These differences more likely caused by the higher uncertainties in abundance measurements and smaller samples for [Fe/H] $<-1$. At the metal-rich regime ([Fe/H] $> 0$), for all alpha elements except calcium, the abundances of field $(0,-10)$ are enhanced compared to field $(0,-5)$. This can be seen most clearly for [$\alpha$/Fe], but is much less certain for [O/Fe]. {We note, however, that there are fewer stars at the high metallicity regime, especially for field $(0,-10)$.}

The lack of vertical alpha abundance gradient in each metallicity component for [Fe/H] $<$ 0 is indicative of fast bulge evolution. Similarly, it has been shown that the high-$\alpha$ disk population (commonly referred to as the thick disk) does not exhibit a vertical [$\alpha$/Fe] gradient~\citep{Ruchti2011,Mikolaitis2014,Duong2018}. 

\subsection{Comparison with GALAH DR2}
\label{subsec:galahcomp}
\begin{figure}
	\centering
	\includegraphics[width=1.01\columnwidth]{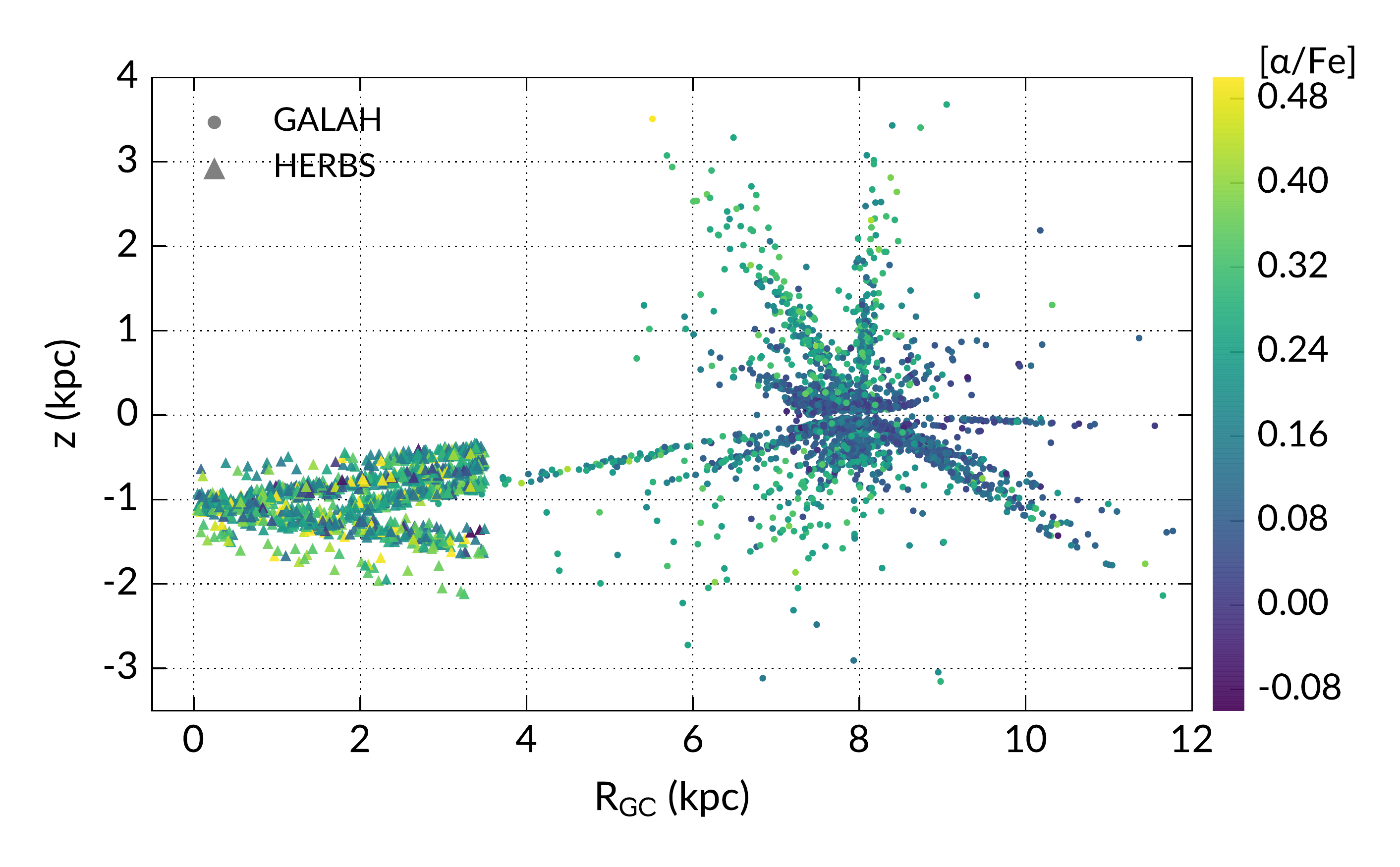}
	\caption{The distribution of HERBS (triangle) and GALAH (circle) giants in the $R_\mathrm{GC}$-$z$ plane. We assumed $R_\odot = 8$ kpc. HERBS stars are located near the Galactic centre ($R_\mathrm{GC} < 4$ kpc) and close to the plane. Most of the GALAH stars are located in the solar neighbourhood due to the survey observing strategy.}
	\label{fig:galah-rz}
\end{figure}

Because this work uses the same lines, atomic data and spectral analysis technique as the GALAH survey, systematic differences and offsets are minimal (see Appendix \ref{A}). Although differences are expected due to our independent reduction and lower S/N, our bulge sample allows for a consistent comparison with disk/halo stars from the GALAH survey DR2~\citep{Buder2018}. To avoid intrinsic offsets in the abundance ratios of different stellar types, we restricted the GALAH sample to approximately same parameters space as that shown in Fig.~\ref{fig:HR}: $T_\mathrm{eff} \approx$ 4000--5000 K and $\log g \approx $ 3.5--1.5 cms$^{-2}$. For a fair comparison, we only used results determined by SME, i.e., the reference results used to train \emph{The Cannon}. The GALAH training set is of high fidelity and signal-to-noise, with mean SNR of $\approx$100 per pixel for the green CCD. Due to the survey observing strategy, most GALAH stars are in the outer disk. Assuming $R_\odot = 8$ kpc, over 90\% of the comparison sample are located at $R_\mathrm{GC} > 7$ kpc and 80\% are in the solar neighbourhood ($7 < R_\mathrm{GC} < 9$ kpc). The $R_\mathrm{GC}$-$z$ distribution of GALAH giants is shown in Fig.~\ref{fig:galah-rz}.

\begin{figure*}
	\centering 
\includegraphics[width=0.5\textwidth]{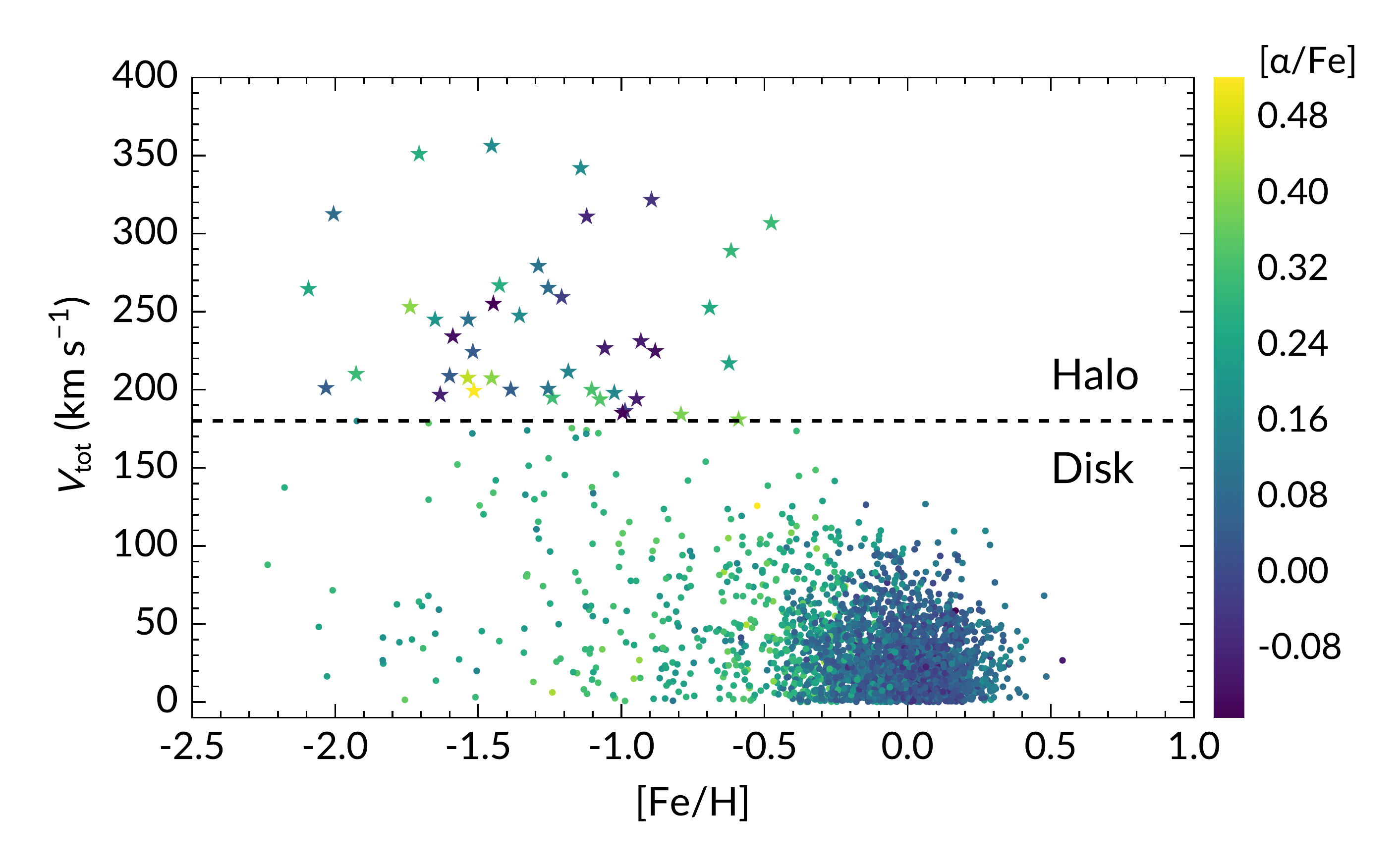}\includegraphics[width=0.5\textwidth]{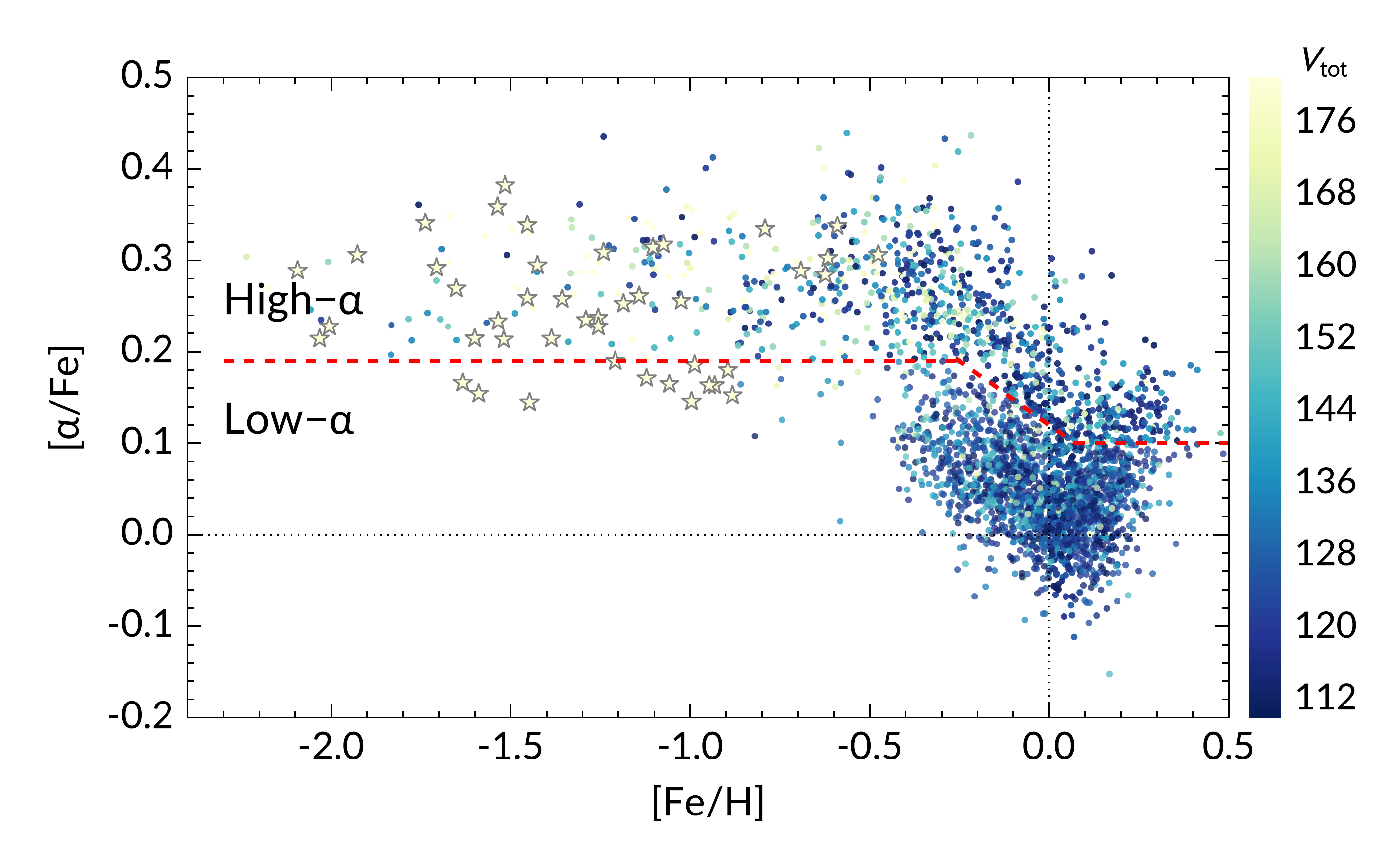}	
\caption{Definitions of stellar populations in the GALAH sample. Left panel: $V_\mathrm{tot}$ as a function of metallicity. The dotted line shows $V_\mathrm{tot} = 180$ km s$^{-1}$, which separates the halo from the disk components. Right panel: The [$\alpha$/Fe] distribution of GALAH disk stars; stars above the dashed line is defined as the high-$\alpha$ population, and below the dashed line is the low-$\alpha$ population. Large stars indicate the halo population ($V_\mathrm{tot} > 180$ km s$^{-1}$).}
	\label{fig:halo-disk-sep}
\end{figure*}

We determined the component membership of the GALAH disk stars based on their weighted average [$\alpha$/Fe], which shows the clearest separation between the disk components. We separated the low and high-\afe~populations of the disk guided by the `gap' in the [$\alpha$/Fe] distribution. The separation is shown in Fig~\ref{fig:halo-disk-sep}. At the metal-poor regime ([Fe/H] $\leq -1$), there is substantial overlap between the metal-weak thick disk and halo in velocities, metallicity and abundances~\citep[e.g.,][]{Reddy2008}. To identify halo stars, we relied on the space velocities $U, V, W$, computed using \emph{Gaia} DR2 parallaxes and proper motions~\citep{Prusti2016,Brown2018}. The solar motion is corrected by adapting $\left(U, V, W\right)_\odot = (11.1, 12.24, 7.25)$ from~\cite{Schonrich2010}. Fig \ref{fig:halo-disk-sep} shows the total velocity $V_\mathrm{tot} \equiv \sqrt{\mathstrut U^2 + V^2 + W^2}$ as a function of [Fe/H]. We designated those with $V_\mathrm{tot} > 180$ km s$^{-1}$ as likely halo stars~\citep[e.g.,][]{Nissen2010}. 

\begin{figure*}
	\includegraphics[width=1\columnwidth]{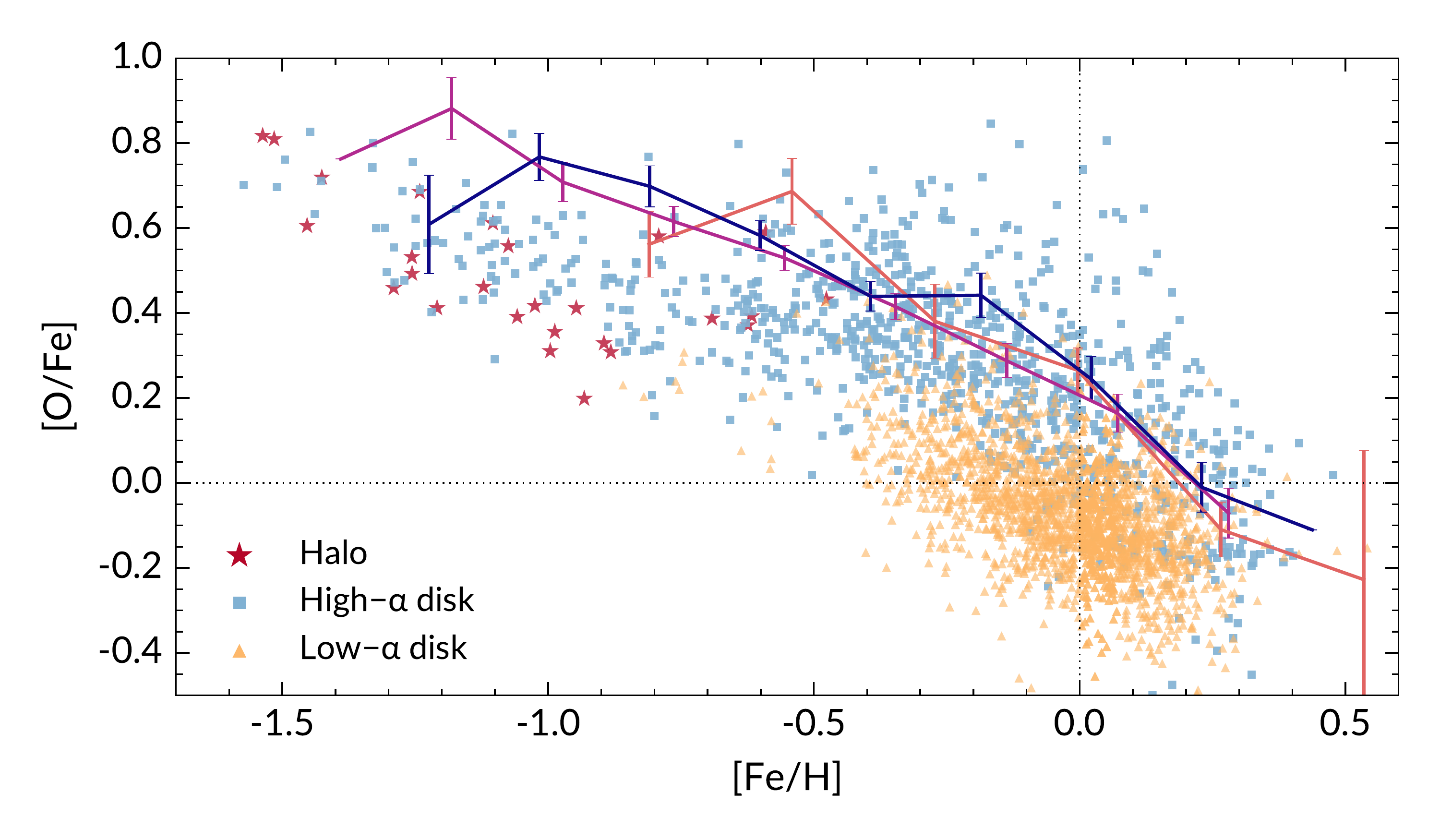}\includegraphics[width=1\columnwidth]{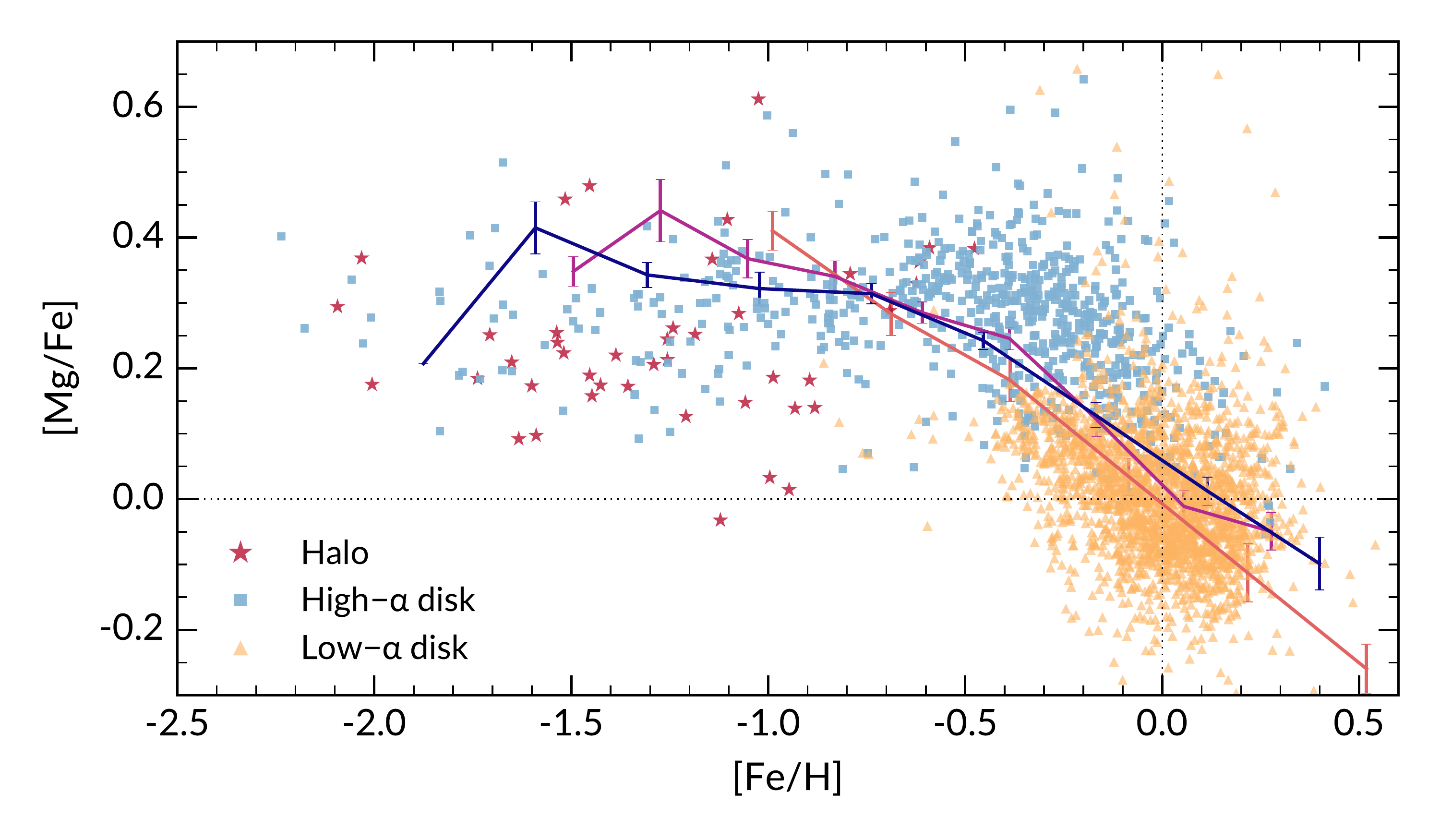}\\
	\includegraphics[width=1\columnwidth]{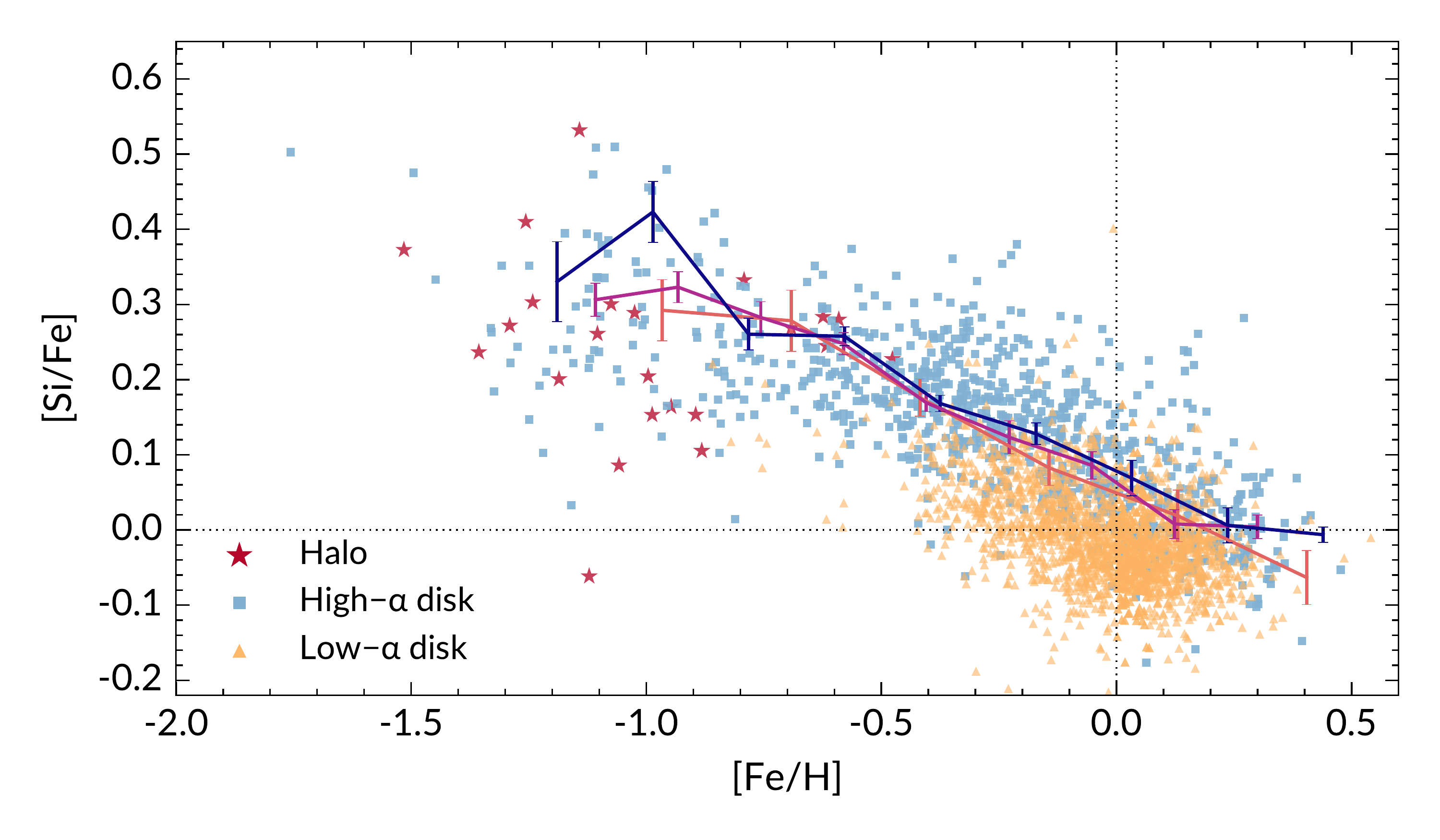}\includegraphics[width=1\columnwidth]{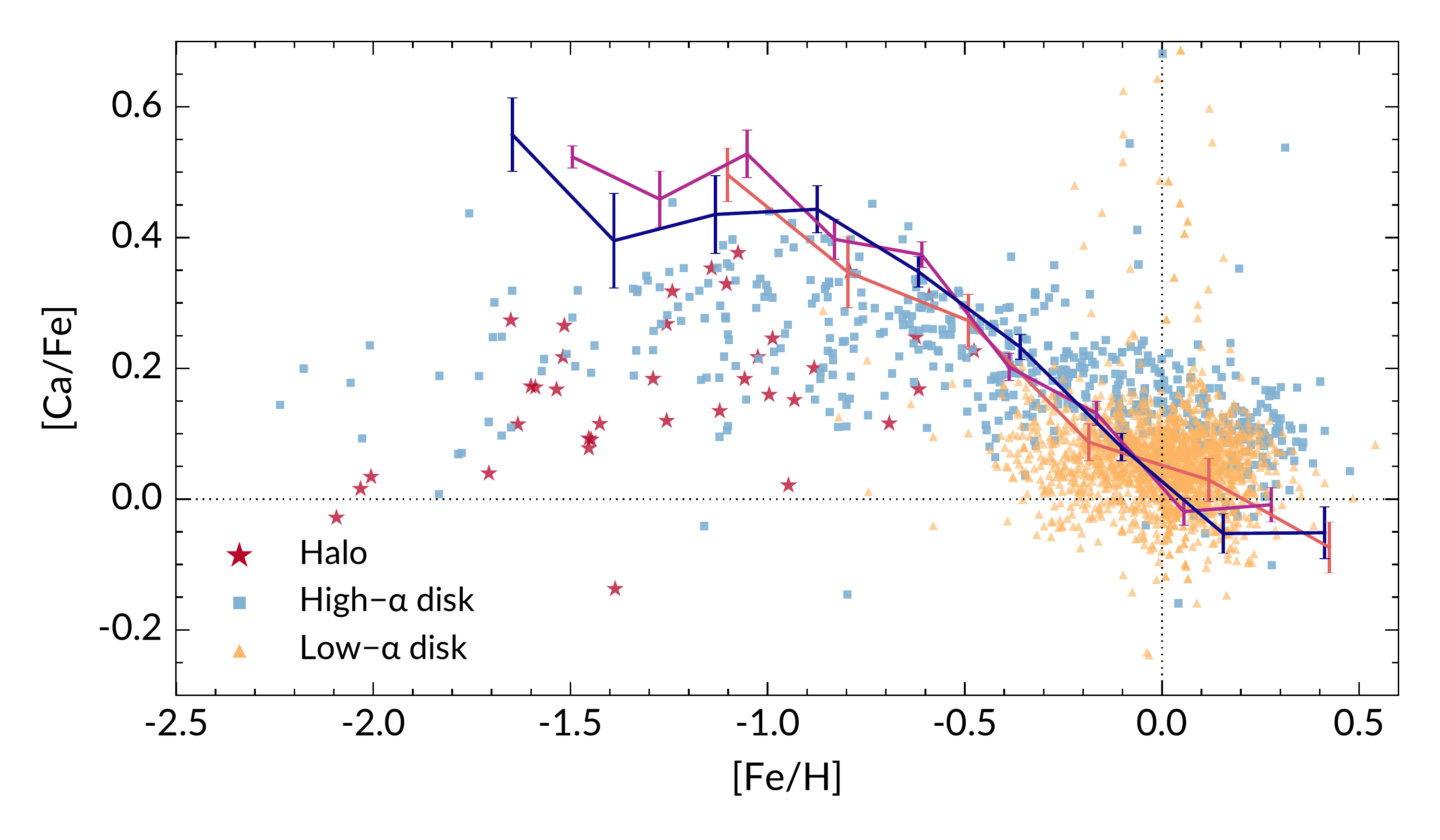}\\
	\includegraphics[width=1\columnwidth]{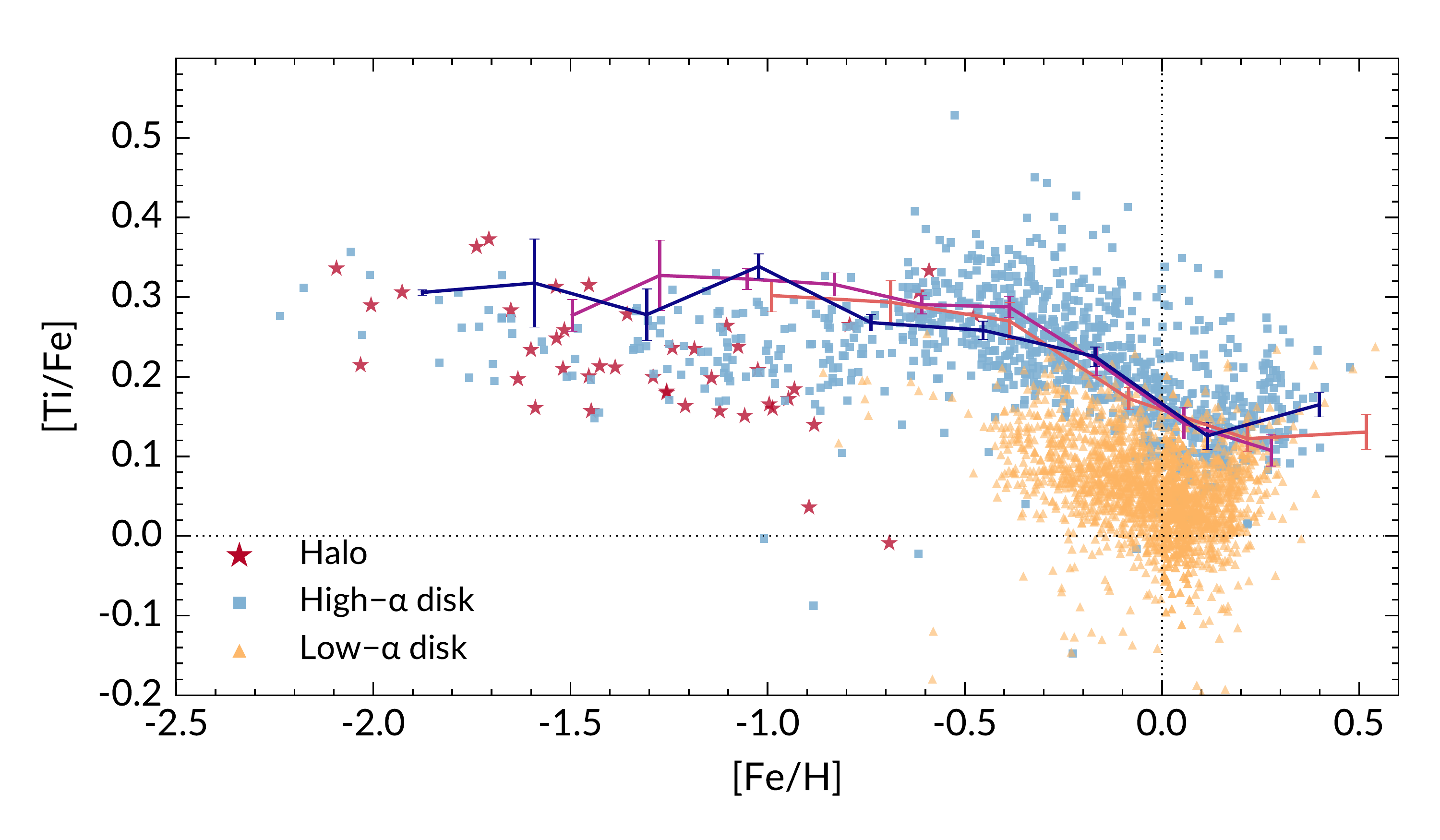}
	\caption{Comparison of the abundance trends in the Galactic bulge and disk/halo. The data points are training set giants from GALAH DR2, separated into the disk and halo components as described in the text. The solid lines are median abundance trends of three bulge fields along the minor axis (colours have the same meaning as in Fig~\ref{fig:abundlat}).}
	\label{fig:galahcomp}
\end{figure*}

Fig \ref{fig:galahcomp} shows the comparison of bulge and disk/halo abundance trends for the five alpha elements and weighted average [$\alpha$/Fe]. Our abundance trends and scale are mostly compatible with GALAH, as we would expect. We note that our [Mg/Fe] trend does not resemble the GALAH trend as do the other elements. This may be due to the different Mg lines used in this study and GALAH (we omitted a magnesium line at 4730 \AA~due to blending). For this reason, we do not show the comparison for [$\alpha$/Fe], as this average would be affected by the systematic difference between our and GALAH [Mg/Fe] ratios. 

Overall the bulge trend follows that of the high-$\alpha$ disk component for [Fe/H] $\gtrsim -0.8$. The bulge abundances remain enhanced compared to the low-$\alpha$ component also at the metal-rich regime. {However, both Mg and Ca abundance ratios show little difference between the low-$\alpha$ disk and bulge, especially at high metallicity. We note the high and low-$\alpha$ disks are not easily separated in the GALAH [Mg/Fe] and [Ca/Fe] distributions, perhaps due to the lower precision of these measurements (see \cite{Buder2018} for details).} {Except for Mg and Ca, the rest of the alpha elements (O, Si, Ti) show behaviours that are in line with the conclusion of many previous works:} that the bulge and high-$\alpha$ disk population shares a similar chemical evolution~\citep{Melendez2008,Alves-Brito2010, Johnson2014, Rojas-Arriagada2017, Jonsson2017, Bensby2017}. There is thus good evidence to support a disk origin for bulge stars with [Fe/H] $> -0.8$. We note that \cite{McWilliam2016} concluded the bulge is enhanced in [Mg/Fe] compared to the thick disk by examining several literature studies, however this comparison may be affected by systematic offsets between bulge RGB and thick disk main-sequence stars. 

The metal-poor bulge population ([Fe/H] $\lesssim -0.8$), however, appears to be enhanced in some alpha elements compared to the thick disk and halo by $\approx 0.1$ dex. This result is less conclusive for [Si/Fe], where the scatter at low metallicity is higher than other elements.

%\begin{table}
%	\caption{The median of disk/halo and bulge at [Fe/H] $\lesssim -0.8$ for each alpha element.}
%	\label{table:metalpoordiff}
%	\begin{tabular}{lll}
%		\hline 
%        Abundance ratio & Bulge median & Disk median\\
%        \hline
%       {[O/Fe]} & 0.73 $\pm$ 0.07 & 0.55 $\pm$ 0.07\\
%        {[Mg/Fe]} & 0.35 $\pm$ 0.04 & 0.25 $\pm$ 0.04\\
%        {[Si/Fe]} & 0.30 $\pm$ 0.03 & 0.20 $\pm$ 0.04\\
%        {[Ca/Fe]} & 0.49 $\pm$ 0.05 & 0.27 $\pm$ 0.06\\
%        {[Ti/Fe]} & 0.33 $\pm$ 0.03 & 0.22 $\pm$ 0.03\\
%        \hline
%	\end{tabular}
%\end{table}

The enhanced alpha abundance ratios suggest a slightly higher star formation rate (SFR) for the metal-poor bulge population, as the initial mass function (IMF) of the bulge has shown to be consistent with that of the local disk~\citep[e.g.,][]{Calamida2015,Wegg2017}. The metal-poor bulge population also has distinctive kinematics signatures:~\cite{Ness2013a} found that stars with [Fe/H] $< -1$ have a different rotation profile to the metal-rich stars, and \cite{Zoccali2017} showed that their metal-poor bulge stars rotate more slowly. This would suggest that the metal-poor population may not share the same disk origin as the more metal-rich populations. 

\section{Conclusion}

In this work, we have successfully obtained stellar parameters and $\alpha$-element abundances for 832 RGB stars at latitudes $b = -5^\circ, -7.5^\circ, -10^\circ$ along the minor axis of the Galactic bulge. The majority of our sample are ARGOS survey stars with pre-determined bulge memberships. ARGOS stars were selected based on metallicity so that we observe higher relative fractions of the metal-rich and metal-poor bulge populations.

According to our analysis of stars in common with ARGOS, the metallicity scale reported by \cite{Ness2013} should be compressed; i.e., we obtain slightly higher [Fe/H] for metal-poor ARGOS stars, and vice versa. Our results suggest that along the minor axis, the spacing between ARGOS MDF component centroids should be $\approx$0.09 dex closer. The effect of this is most apparent in field $(0,-7.5)$, where primary ARGOS components A and B become almost indistinguishable. However, the ARGOS components remain distinct (given measured ARGOS dispersions) for fields $(0,-5)$ and $(0,-10)$.

The optical wavelength range and resolving power of the HERMES spectrograph allowed us to measure chemical abundances for up to 28 elements, including the alpha elements O, Mg, Si, Ca and Ti. In general, we find that the [X/Fe] vs [Fe/H] trends of these elements follow that of previous works at similar resolving power. We also observe similar trends to the microlensed dwarfs sample from \cite{Bensby2017}, which was observed at much higher resolving power but is limited to less than 100 stars. Within the scatter of the datasets, we confirm that there are no significant systematic differences between the bulge giants and microlensed dwarfs, except for [Ca/Fe] from [Fe/H] $< -0.5$, where our median [Ca/Fe] is higher by up to 0.2 dex. In addition, our [Mg/Fe] values decrease as a function of [Fe/H] and do not flatten at super-solar metallicity. We find that the mean value of [$\alpha$/Fe] increases with increasing distance from the plane, which is expected as the metal-poor component dominates at high latitudes. We also find that the [$\alpha$/Fe] dispersion is smaller at higher latitudes. 

Our metallicity coverage allowed us to assess the vertical variation in alpha abundances of the different bulge metallicity components. Within uncertainties, the abundance ratios remain uniform with height for most metallicity bins. At the metal-rich regime ([Fe/H] $>$ 0), there is evidence of enhanced alpha-abundances in field $(0,-10)$, which is most conclusive for [$\alpha$/Fe] (weighted average of Mg, Ca, Si, Ti). However, this conclusion is more uncertain for individual elements, and does not seem to hold true for [Ca/Fe]. The bulge abundance trends appear to follow that of the high-$\alpha$ disk population, and are enhanced compared to the low-$\alpha$ disk population at super solar metallicities. However, the more metal-poor bulge population ([Fe/H] $\lesssim -0.8$) is enhanced compared to thick disk and halo stars at the same metallicity. 

The lack of vertical abundance variation for different metallicity components and abundance trends similar to the high-$\alpha$, or thick disk population, both point to fast chemical enrichment in the bulge~\citep[e.g.,][]{Friaca2017}. Furthermore, the metal-poor bulge population may have experienced a different evolution, as we observe that it is enhanced in alpha abundances compared to the high-$\alpha$ disk population. This may be compatible with previous findings that the metal-poor population has distinct kinematics compared to the metal-rich population, and indicates that the bulge does not just consist of stars originating from the disk. We further explore the chemical evolution of the bulge and its connection to the disk in the next paper of this series, which will focus on the abundances light, iron peak and heavy elements. 

\section*{Acknowledgements}

We thank the anonymous referee for helpful comments that improved this manuscript. LD, MA and KCF acknowledge funding from the Australian Research Council (projects FL110100012 and DP160103747). LD gratefully acknowledges a scholarship from Zonta International District 24. DMN was supported by the Allan C. and Dorothy H. Davis Fellowship. LMH was supported by the project grant `The New Milky Way' from the Knut and Alice Wallenberg foundation. MA's work was conducted as part of the research by Australian Research Council Centre of Excellence for All Sky Astrophysics in 3 Dimensions (ASTRO 3D), through project number CE170100013. Part of this research was conducted at the Munich Institute for Astro- and Particle Physics (MIAPP) of the DFG cluster of excellence ``Origin and Structure of the Universe''. This publication makes use of data products from the Two Micron All Sky Survey, which is a joint project of the University of Massachusetts and the Infrared Processing and Analysis Center/California Institute of Technology, funded by the National Aeronautics and Space Administration and the National Science Foundation. We acknowledge the use of data from the European Space Agency (ESA) mission
{\it Gaia} (\url{https://www.cosmos.esa.int/gaia}), processed by the {\it Gaia} Data Processing and Analysis Consortium (DPAC,
\url{https://www.cosmos.esa.int/web/gaia/dpac/consortium}). Funding for the DPAC has been provided by national institutions, in particular the institutions participating in the {\it Gaia} Multilateral Agreement. The GALAH survey is based on observations made at the Australian Astronomical Observatory, under programmes A/2013B/13, A/2014A/25, A/2015A/19, A/2017A/18. We acknowledge the traditional owners of the land on which the AAT stands, the Gamilaraay people, and pay our respects to elders past and present. 

%%%%%%%%%%%%%%%%%%%% REFERENCES %%%%%%%%%%%%%%%%%%
% The best way to enter references is to use BibTeX:

\bibliographystyle{mnras}
\bibliography{Ref2} % if your bibtex file is called example.bib

\appendix

\section{Spectroscopic analysis tests}
\label{A}
\subsection{Benchmark stellar parameters comparison}
\label{sec:benchmark}
\begin{figure}
	\centering
	\includegraphics[width=1.\columnwidth]{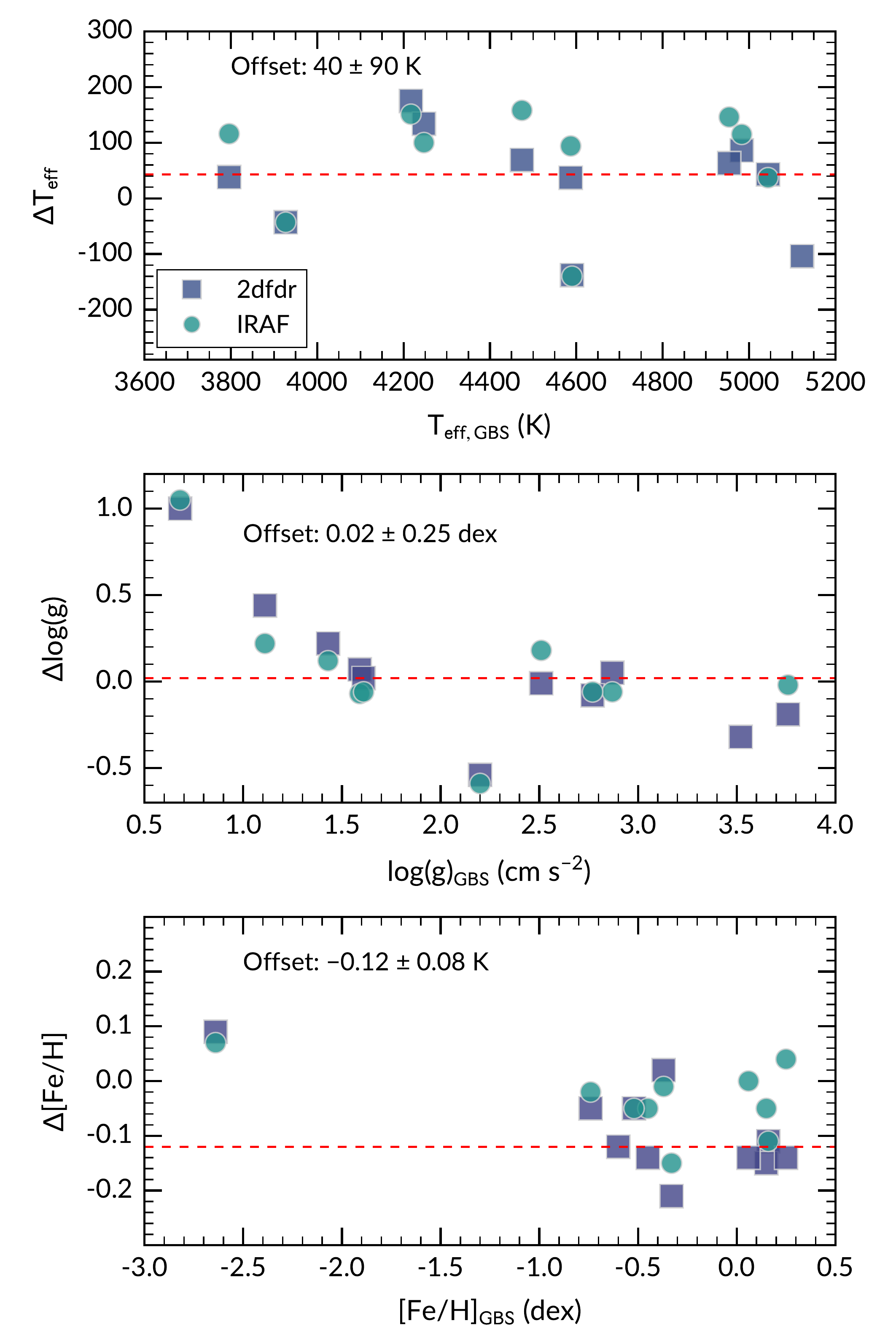}
	\caption{Comparison of SME-derived parameters with fundamental $T_\mathrm{eff}$, $\log g$~\citep{Heiter2015} and $Gaia$ ESO-derived [Fe/H] (from high resolution UVES spectra,~\citealt{Jofre2014}) for benchmark stars. The squares indicate 2dfdr-based reductions and the circles are results from IRAF-based reductions. The differences are shown as (SME $-$ GBS); the red line indicates biases for 2dfdr-based reductions \emph{only}. The outlier in $\log g$ is the M-giant $\alpha$Ceti.}
	\label{fig:paramvalid}
\end{figure}
\begin{figure}
	\centering
	\includegraphics[width=1\columnwidth]{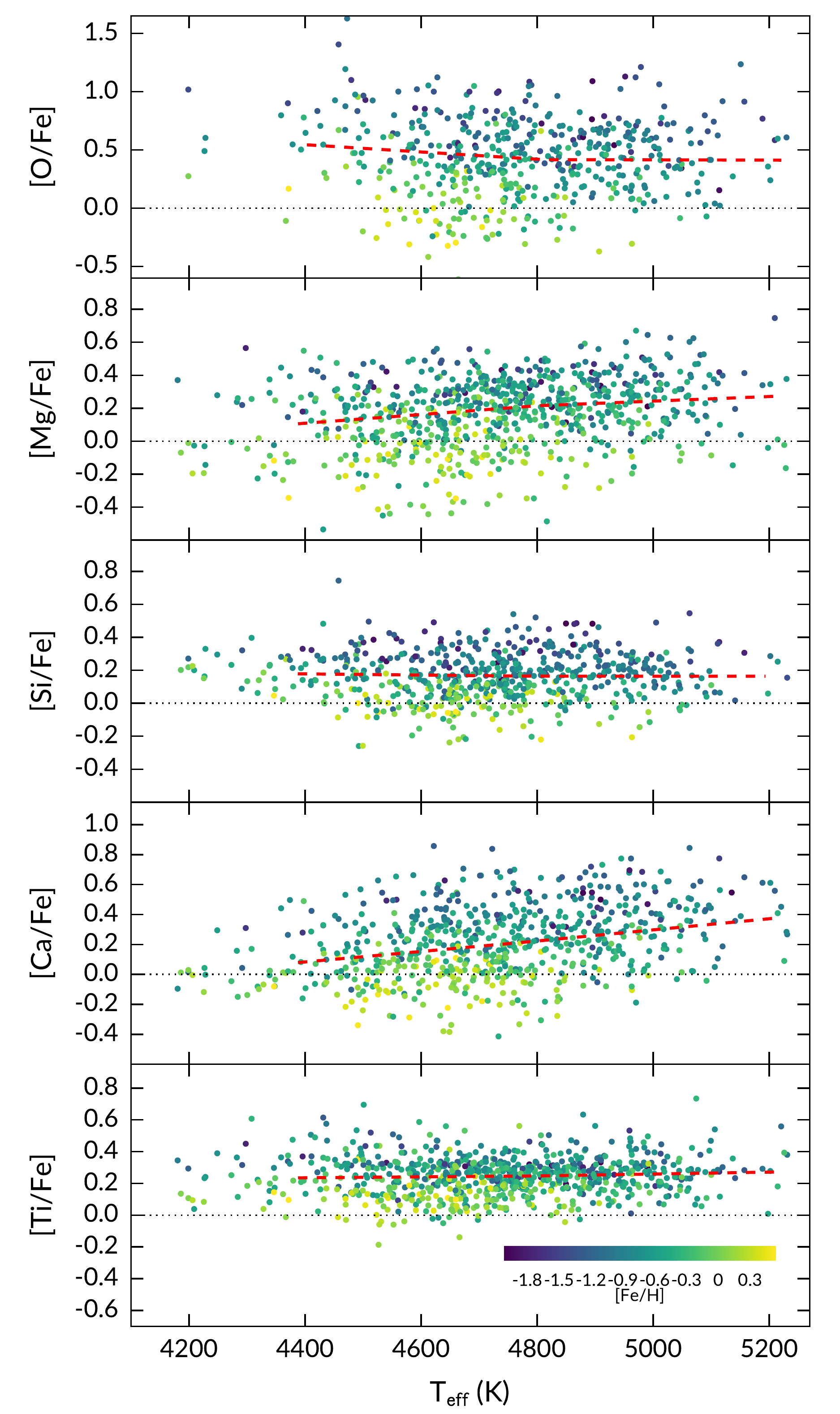}
	\caption{Abundance-temperature trends for the alpha elements; the red dashed lines are running medians over four metallicity bins. We do not see obvious correlations in [O/Fe], [Si/Fe] and [Ti/Fe]. Similar linear trends are observed for [Mg/Fe] and [Ca/Fe].}
	\label{fig:abundteff}
\end{figure}
To estimate the accuracy of our stellar parameters, we analysed giants and subgiants in the $Gaia$ benchmark stars (GBS) sample {(giants only)} using the reduction and parameter optimisation pipeline described above. For this purpose we used archive HERMES benchmark observations that were taken for the GALAH survey. To gauge the effects of the difference in reduction methods between our survey (2dfdr-based) and the GALAH survey (IRAF-based), we also compared our benchmark results with that obtained from spectra reduced with the GALAH reduction pipeline. The results are shown in Fig. \ref{fig:paramvalid}. We note that the GALAH spectrum of the benchmark giant $\epsilon$For was not available, but this star is included in the 2dfdr comparison sample. We performed our analysis assuming $\mathcal{R}$ = 28 000. For the IRAF benchmark reductions, adopting either this constant spectral resolution or the resolution map from~\citep{Kos2017} returned near identical results, with temperature differences $\leq 30$ K, metallicity differences $\leq 0.04$ dex and surface gravity differences $\leq 0.1$ dex, all within expected uncertainties. 

In general, the differences between the two reduction methods are not significant. For the 2dfdr sample, we observe biases\footnote{The biases are defined as the median of the difference (SME $-$ benchmark).} and standard deviations for $T_\mathrm{eff}$: $40 \pm 90$ K; $\log g$: $0.02 \pm 0.25$ dex and [Fe/H]: $-0.12 \pm 0.08$ dex. The IRAF biases and standard deviations for $T_\mathrm{eff}$, $\log g$ and [Fe/H] are: $100 \pm 90$ K; $-0.06 \pm 0.22$ dex and $-0.05 \pm 0.06$ dex, respectively. While both sets of results are fairly accurate, spectra from the GALAH reduction pipeline produce slightly more precise $\log g$ and [Fe/H], perhaps due to the tilted PSF correction that increases the resolution and signal-to-noise towards the CCD corners. However, GALAH-reduced spectra return even higher effective temperature for giants compared to our reductions, which already overestimates $T_\mathrm{eff}$ by 40 K compared to reference values.

For the GALAH reductions, the biases of this giants-only sample is different to that of the full giants and dwarfs benchmark sample analysed by \cite{Sharma2018}. Before bias correction, \cite{Sharma2018} quoted biases in $T_\mathrm{eff}$: $23 \pm 112$ K; $\log g$: $-0.15 \pm 0.22$ dex and [Fe/H]: $-0.12 \pm 0.1$ dex. Here, we find that neither $\log g$ nor [Fe/H] show a strong offset, and the metallicity precision is higher compared to the full sample. 

We derived particularly discrepant surface gravity results for the M-giant $\alpha$Ceti. Compared to the benchmark value from \cite{Heiter2015}, this star has $\Delta \log g$ = 1 dex, for both 2dfdr and IRAF reductions. The cause of this discrepancy is unclear, seeing as the M-giant $\alpha$Tau with similar parameters does not show this large deviation. This may be a concern that surface gravity determination for $\log g <$ 1 dex is challenging with HERMES spectra. The outlier $\alpha$Ceti has been excluded from $\log g$ standard deviation calculations. {Another notable discrepant point is the cool giant HD107328, with $\Delta \log g = -0.56$. \cite{Buder2018} showed that a result close to the GBS value can only be produced if $Gaia$ DR1 parallax is used. This indicates there are some difficulties in analysing line-rich stars with our spectroscopic method and wavelength region.}

Our comparison here shows that 2dfdr-reduced spectra perform just as well as IRAF-reduced spectra, however the IRAF reductions return higher precision for [Fe/H] and $\log g$. We do not see significant biases in effective temperature or surface gravity in our results, and thus do make corrections to these parameters. However, the metallicity is underestimated by our pipeline by $-0.12$ dex. Similar to GALAH, we corrected for this metallicity bias by adding +0.1 dex to all of our [Fe/H] values. 

\subsection{Abundance trends with temperature}
Fig. \ref{fig:abundteff} shows the abundance-temperature trends for each $\alpha$-element. We do not observe correlations for oxygen, silicon and titanium. [Mg/Fe] and [Ca/Fe] both show positive linear trends with respect to temperature. For calcium, this could be due to non-LTE effects, however the reason for the [Mg/Fe]-$T_\mathrm{eff}$ correlation is not clear. The non-LTE magnesium abundances of M67 giants observed by GALAH also the same trend with temperature (see~\citealt{Gao2018}, their Fig. 7). 
%Atomic data: all lines for stellar parameter determination, all lines for abundance determination (1 line for O, 2 lines for Mg,  2 lines for Si,  2 lines for Ca, 10+lines for Ti)

\section{Data tables}
\label{B} 
The catalogue containing stellar parameters, abundance ratios and their uncertainties is provided as online supporting material, accessible through the publisher website. The contents of the catalogue is described in Table~\ref{table:data}. We also provide the list of lines used for stellar parameters and abundance analysis in Tables~\ref{table:atomic} and~\ref{table:atomic2}. The atomic data and lines used for stellar parameters analysis are the same as that of the GALAH survey~\citep{Buder2018}. Except for Ti, we did not use all of the lines available in the GALAH linelist for abundance analysis. 
\begin{table}
	\caption{Description of the data catalogue. The uncertainties in abundance ratios are $\chi^2$ fitting errors; the uncertainties in stellar parameters are $\chi^2$ fitting errors and standard deviations from Fig. \ref{fig:paramvalid} added in quadrature.}
	\label{table:data}
	\begin{tabular}{lll}
		\hline 
        Column & Name & Description\\
        \hline
        {[1]} & 2MASS ID & The 2MASS identifier of the star\\
        {[2]} & RAJ2000 & The right ascension at epoch J2000 (degrees)\\
        {[3]} & DECJ2000 & The declination at epoch J2000 (degrees)\\
        {[4]} & $T_\mathrm{eff}$ & Effective temperature (K)\\
        {[5]} & $\sigma_\mathrm{T_{eff}}$ & Uncertainty in effective temperature (K)\\
        {[6]} & $\log g$ & Surface gravity (cm s$^{-2}$)\\
        {[7]} & $\sigma_{\log g}$ & Uncertainty in surface gravity (cm s$^{-2}$)\\
        {[8]} & [Fe/H] & Metallicity\\
		{[9]} & $\sigma_\mathrm{[Fe/H]}$ & Uncertainty in metallicity\\
        {[10]} & [O/Fe] & Abundance ratio for O\\
        {[11]} & $\sigma_\mathrm{[O/Fe]}$ & Uncertainty in [O/Fe]\\
        {[12]} & [X/Fe] & Same as [10], but for Mg\\
        {[13]} & $\sigma_\mathrm{[X/Fe]}$ & Same as [11], but for Mg\\
        {[14]} & [X/Fe] & Same as [10], but for Si\\
        {[15]} & $\sigma_\mathrm{[X/Fe]}$ & Same as [11], but for Si\\
        {[16]} & [X/Fe] & Same as [10], but for Ca\\
        {[17]} & $\sigma_\mathrm{[X/Fe]}$ & Same as [11], but for Ca\\
        {[18]} & [X/Fe] & Same as [10], but for Ti\\
        {[19]} & $\sigma_\mathrm{[X/Fe]}$ & Same as [11], but for Ti\\
        {[20]} & ARGOS $T_\mathrm{eff}$ & ARGOS effective temperature (K)\\
        {[21]} & ARGOS $\log g$ & ARGOS surface gravity (cm s$^{-2}$)\\
        {[22]} & ARGOS [Fe/H] & ARGOS metallicity\\
        \hline
	\end{tabular}
\end{table}
\begin{table*}
	\caption{Line data used for stellar parameters determination, common to this work and GALAH DR2.}
	\label{table:atomic}
	\begin{tabular}{ccccccccc}
		\hline 
		Species & Wavelength (\AA) & $\log \left(gf\right)$ & Excitation potential (eV)& &Species & Wavelength (\AA) & $\log \left(gf\right)$ & Excitation potential (eV)\\
		\hline
\ion{Sc}{i}	&	4743.8300	&	0.422	&	1.448&	&	\ion{Fe}{i}	&	5661.3447	&	-1.756	&	4.284\\
\ion{Sc}{i}	&	4753.1610	&	-1.659	&	0.000&	&	\ion{Fe}{i}	&	5662.5161	&	-0.447	&	4.178\\
\ion{Sc}{i}	&	5671.8163	&	-0.290	&	1.448&	&	\ion{Fe}{i}	&	5679.0229	&	-0.820	&	4.652\\
\ion{Sc}{i}	&	5686.8386	&	-0.133	&	1.440&	&	\ion{Fe}{i}	&	5680.2404	&	-2.480	&	4.186\\
\ion{Sc}{i}	&	5717.3070	&	-0.532	&	1.440&	&	\ion{Fe}{i}	&	5696.0892	&	-1.720	&	4.549\\
\ion{Sc}{i}	&	5724.1070	&	-0.661	&	1.433&	&	\ion{Fe}{i}	&	5701.5442	&	-2.193	&	2.559\\
\ion{Sc}{ii}	&	5657.8960	&	-0.603	&	1.507&	&	\ion{Fe}{i}	&	5705.4642	&	-1.355	&	4.301\\
\ion{Sc}{ii}	&	5667.1490	&	-1.309	&	1.500&	&	\ion{Fe}{i}	&	5731.7618	&	-1.200	&	4.256\\
\ion{Sc}{ii}	&	5684.2020	&	-1.074	&	1.507&	&	\ion{Fe}{i}	&	5732.2960	&	-1.460	&	4.991\\
\ion{Sc}{ii}	&	6604.6010	&	-1.309	&	1.357&	&	\ion{Fe}{i}	&	5741.8477	&	-1.672	&	4.256\\
\ion{Sc}{ii}	&	5669.0420	&	-1.200	&	1.500&	&	\ion{Fe}{i}	&	5775.0805	&	-1.080	&	4.220\\
\ion{Ti}{i}	&	4758.1178	&	0.510	&	2.249&	&	\ion{Fe}{i}	&	5778.4533	&	-3.430	&	2.588\\
\ion{Ti}{i}	&	4759.2697	&	0.590	&	2.256&	&	\ion{Fe}{i}	&	5806.7249	&	-0.950	&	4.608\\
\ion{Ti}{i}	&	4778.2547	&	-0.350	&	2.236&	&	\ion{Fe}{i}	&	5809.2174	&	-1.740	&	3.884\\
\ion{Ti}{i}	&	4781.7106	&	-1.950	&	0.848&	&	\ion{Fe}{i}	&	5811.9144	&	-2.330	&	4.143\\
\ion{Ti}{i}	&	4797.9757	&	-0.630	&	2.334&	&	\ion{Fe}{i}	&	5814.8071	&	-1.870	&	4.283\\
\ion{Ti}{i}	&	4801.9016	&	-3.060	&	0.818&	&	\ion{Fe}{i}	&	5849.6833	&	-2.890	&	3.695\\
\ion{Ti}{i}	&	4820.4094	&	-0.380	&	1.503&	&	\ion{Fe}{i}	&	5853.1483	&	-5.180	&	1.485\\
\ion{Ti}{i}	&	5689.4600	&	-0.360	&	2.297&	&	\ion{Fe}{i}	&	5855.0758	&	-1.478	&	4.608\\
\ion{Ti}{i}	&	5716.4500	&	-0.720	&	2.297&	&	\ion{Fe}{i}	&	5858.7780	&	-2.160	&	4.220\\
\ion{Ti}{i}	&	5720.4359	&	-0.900	&	2.292&	&	\ion{Fe}{i}	&	6481.8698	&	-2.981	&	2.279\\
\ion{Ti}{i}	&	5739.4690	&	-0.610	&	2.249&	&	\ion{Fe}{i}	&	6494.9804	&	-1.268	&	2.404\\
\ion{Ti}{i}	&	5866.4513	&	-0.790	&	1.067&	&	\ion{Fe}{i}	&	6498.9383	&	-4.687	&	0.958\\
\ion{Ti}{i}	&	6716.6660	&	-1.370	&	2.488&	&	\ion{Fe}{i}	&	6546.2381	&	-1.536	&	2.759\\
\ion{Ti}{ii}	&	4719.5109	&	-3.320	&	1.243&	&	\ion{Fe}{i}	&	6592.9124	&	-1.473	&	2.728\\
\ion{Ti}{ii}	&	4764.5247	&	-2.690	&	1.237&	&	\ion{Fe}{i}	&	6593.8695	&	-2.420	&	2.433\\
\ion{Ti}{ii}	&	4798.5313	&	-2.660	&	1.080&	&	\ion{Fe}{i}	&	6597.5592	&	-0.970	&	4.795\\
\ion{Ti}{ii}	&	4849.1678	&	-2.960	&	1.131&	&	\ion{Fe}{i}	&	6609.1097	&	-2.691	&	2.559\\
\ion{Ti}{ii}	&	4865.6104	&	-2.700	&	1.116&	&	\ion{Fe}{i}	&	6627.5438	&	-1.590	&	4.549\\
\ion{Ti}{ii}	&	4874.0094	&	-0.860	&	3.095&	&	\ion{Fe}{i}	&	6648.0796	&	-5.918	&	1.011\\
\ion{Fe}{i}	&	4788.7566	&	-1.763	&	3.237&	&	\ion{Fe}{i}	&	6677.9851	&	-1.418	&	2.692\\
\ion{Fe}{i}	&	4793.9614	&	-3.430	&	3.047&	&	\ion{Fe}{i}	&	6699.1413	&	-2.101	&	4.593\\
\ion{Fe}{i}	&	4794.3541	&	-3.950	&	2.424&	&	\ion{Fe}{i}	&	6703.5660	&	-3.060	&	2.759\\
\ion{Fe}{i}	&	4802.8797	&	-1.510	&	3.642&	&	\ion{Fe}{i}	&	6713.7425	&	-1.500	&	4.795\\
\ion{Fe}{i}	&	4808.1478	&	-2.690	&	3.251&	&	\ion{Fe}{i}	&	6725.3558	&	-2.100	&	4.103\\
\ion{Fe}{i}	&	4875.8770	&	-1.900	&	3.332&	&	\ion{Fe}{i}	&	6733.1503	&	-1.480	&	4.638\\
\ion{Fe}{i}	&	4890.7551	&	-0.386	&	2.876&	&	\ion{Fe}{ii}	&	4720.1386	&	-4.480	&	3.197\\
\ion{Fe}{i}	&	4891.4921	&	-0.111	&	2.851&	&	\ion{Fe}{ii}	&	4731.4476	&	-3.100	&	2.891\\
\ion{Fe}{i}	&	5651.4689	&	-1.900	&	4.473&	&	\ion{Fe}{ii}	&	4833.1916	&	-5.110	&	2.657\\
\ion{Fe}{i}	&	5652.3176	&	-1.850	&	4.260 &	\\
\hline
\end{tabular}
\end{table*}

\begin{table*}
	\caption{Line data used for abundance determination. The same Ti lines in table~\ref{table:atomic} were used to determine [Ti/Fe].}
	\label{table:atomic2}
	\begin{tabular}{ccccccccc}
		\hline 
		Species & Wavelength (\AA) & $\log \left(gf\right)$ & Excitation potential (eV)& &Species & Wavelength (\AA) & $\log \left(gf\right)$ & Excitation potential (eV)\\
		\hline
\ion{O}{i} 		& 	7771.9440	& 	0.369	& 	9.146 && \ion{Si}{i}	&	5793.0726	&	-1.963	&	4.930 \\
\ion{Mg}{i}	&	5711.0880	&	-1.724	&	4.346	& & \ion{Si}{i} & 	5701.1040	& 	-1.953	& 	4.930\\
\ion{Mg}{i}	&	7691.5500	&	-0.783	&	5.753 && \ion{Ca}{i} & 	5867.5620	& 	-1.570	& 	2.933\\
\ion{Si}{i}	&	5690.4250	&	-1.773	&	4.930 && \ion{Ca}{i}	&	6499.6500	&	-0.818	& 2.523\\ 
\hline
\end{tabular}
\end{table*}
%%%%%%%%%%%%%%%%%%%%%%%%%%%%%%%%%%%%%%%%%%%%%%%%%%
% Don't change these lines
\bsp	% typesetting comment
\label{lastpage}
\end{document}